\documentclass[twocolumn]{aastex62}		         

\usepackage{mathtools}
\usepackage{float}
\usepackage[]{natbib}
\usepackage{graphicx}	
\graphicspath{{images/}}
\usepackage{amsmath}
\newcommand{\redtext}[1]{\textcolor{black}{#1}}

\usepackage{hyperref}
\usepackage{verbatim}
\usepackage{multirow}

\newcommand{\swift}{\textit{Swift}}
\newcommand{\galex}{\textit{GALEX}}

\newcommand{\swim}{SwiM catalog}
\newcounter{species} 
\def\ion#1#2{\hbox{\setcounter{species}{#2}#1\,{\scriptsize\Roman{species}}\relax}}

\def\lsim{\lower0.3em\hbox{$\,\buildrel <\over\sim\,$}}
\def\gsim{\lower0.3em\hbox{$\,\buildrel >\over\sim\,$}}

\usepackage[normalem]{ulem}
\def\crossout#1{}
\def\crossout#1{\sout{#1}}

\begin{document}

\title{\large \textit{Swift}/UVOT+MaNGA (SwiM) Value-added Catalog}
\correspondingauthor{Nikhil Ajgaonkar}
\email{naj222@g.uky.edu}

\author[0000-0001-8440-3613]{Mallory Molina} 
\affil{Department of Astronomy and Astrophysics and Institute for Gravitation and the Cosmos, The Pennsylvania State University, 525 Davey Lab, University Park, PA 16803, USA}
\affil{eXtreme Gravity Institute, Department of Physics, Montana State University, Bozeman, MT 59715, USA}
\author{Nikhil Ajgaonkar}
\affil{Department of Physics and Astronomy, University of Kentucky, 505 Rose St., Lexington, KY 40506-0057, USA}

\author[0000-0003-1025-1711]{Renbin Yan}
\affil{Department of Physics and Astronomy, University of Kentucky, 505 Rose St., Lexington, KY 40506-0057, USA}

\author{Robin Ciardullo}
\affil{Department of Astronomy and Astrophysics and Institute for Gravitation and the Cosmos, The Pennsylvania State University, 525 Davey Lab, University Park, PA 16803, USA}

\author[0000-0001-6842-2371]{Caryl Gronwall}
\affil{Department of Astronomy and Astrophysics and Institute for Gravitation and the Cosmos, The Pennsylvania State University, 525 Davey Lab, University Park, PA 16803, USA}

\author[0000-0002-3719-940X]{Michael Eracleous}
\affil{Department of Astronomy and Astrophysics and Institute for Gravitation and the Cosmos, The Pennsylvania State University, 525 Davey Lab, University Park, PA 16803, USA}

\author{Xihan Ji}
\affil{Department of Physics and Astronomy, University of Kentucky, 505 Rose St., Lexington, KY 40506-0057, USA}

\author[0000-0003-1641-6222]{Michael R. Blanton}
\affil{Center for Cosmology and Particle Physics, Department of Physics, New York University}

\begin{abstract}
We introduce the \swift/UVOT+MaNGA (SwiM) value added catalog, which comprises 150 galaxies that have both SDSS/MaNGA integral field spectroscopy and archival \textit{Swift}/UVOT near-UV (NUV) images. The similar angular resolution between the three \swift/UVOT NUV images and the MaNGA maps allows for a high-resolution comparison of optical and NUV indicators of star formation, crucial for constraining quenching and attenuation in the local universe. The UVOT NUV images, SDSS images, and MaNGA emission line and spectral index maps have all been spatially matched and re-projected to match the point spread function and pixel sampling of the \textit{Swift}/UVOT uvw2 images, and are presented in the same coordinate system for each galaxy. The spectral index maps use the definition first adopted by \citet{Burstein1984}, which makes it more convenient for users to compute spectral indices when binning the maps. Spatial covariance is properly taken into account in propagating the uncertainties. We also provide a catalog that includes PSF-matched aperture photometry in the SDSS optical and Swift NUV bands. In an earlier, companion paper \citep{molina2020} we used a subset of these galaxies to explore the attenuation laws of kiloparsec-sized star forming regions. The catalog, maps for each galaxy, and the associated data models, are publicly released on the SDSS website\footnote{https://data.sdss.org/sas/dr16/manga/swim/v3.1/}.
\end{abstract}

\keywords{galaxies: general -- astronomy data analysis -- catalogs -- photometry: Sloan -- photometry: ultraviolet -- spectroscopy}

\section{Introduction}
\label{sec:intro}
The growth and quenching of star formation within a galaxy 
are central elements of galaxy evolution. In order to fully understand this process, the appropriate methodological tools and physical models must be in place, including an accurate attenuation law and star formation quenching models. The former is dictated by the intrinsic properties of the regions studied, such as the star formation rate (SFR), stellar mass-specific SFR (log[SFR/M$_*$]) and the chosen sightline, all of which change across the face of the galaxy \citep[e.g.,][]{Charlot2000,Calzetti2000,Wild2011,Xiao2012,Battisti2016,Salim2018}. Similarly, the physical mechanisms that drive the quenching of star formation could be constrained by its spatial progression within a galaxy. For example, both morphological quenching and feedback from active galactic nuclei (AGN) produce inside-out quenching \citep[with different timescales;][]{Martig2009,Springel2005}, while mechanisms such as ram-pressure stripping \citep{Steinhauser2016} would produce outside-in quenching. 

Therefore, spatially resolved measurements of the recent star formation history (SFH) across the faces of galaxies are necessary to study quenching in galaxies. The near ultra-violet (NUV) band and the nebular H$\alpha$ recombination line are robust SFR indicators that probe different timescales \citep[$\sim100$~Myr for NUV, and $\sim10$~Myr for H$\alpha$, see][for details]{Kennicutt1998,Kennicutt2012,Calzetti2013}. Combined with other stellar continuum features, they can provide constraints on the spatial progression of quenching. 
They are also crucial to understanding the relevant attenuation law in star forming regions, including the strength and contribution of the 2175~\AA\ bump \citep[e.g.,][]{Calzetti2000, Battisti2016,Battisti2017,molina2020}. 

The high spatial resolution and narrow NUV filters required for the study of the quenching of star formation are not attainable with the single broad NUV filter and 5\arcsec\ angular resolution of the \textit{Galaxy Evolution Explorer} \citep[GALEX; ][]{Martin2005}.
Therefore, we constructed a catalog of galaxies with Sloan Digital Sky Survey IV (SDSS-IV) Mapping Nearby Galaxies at Apache Point Observatory  \citep[MaNGA;][]{Bundy2015,Yan2016,Blanton2017} optical integral field unit (IFU) spectroscopy and archival \textit{Swift} Ultraviolet Optical Telescope \citep[UVOT;][]{Roming2005} NUV images. The similar angular resolution ($\sim2.\!\!^{\prime\prime}5$) of \textit{Swift}/UVOT and SDSS-IV/MaNGA provides a view of nearby galaxies in both the NUV and optical bands at a spatial resolution of order 1~kpc.

The combination of the spatially-matched NUV images and optical IFU maps 
creates a powerful dataset that can address a number of astrophysical questions. We have used a subset of the galaxies in this catalog in \citet{molina2020}\footnote{\citet{molina2020} used a previous version of the data reduction presented here, which does not include errors due to covariance or matching the \swift/UVOT images to the resolution of the uvw2 filter. Additionally, the D$_n$(4000) error bars in that version were overestimated by including a factor of 1.98 for calibration uncertainties instead of 1.4. However, none of these issues affected the results presented in that work.} to explore the relevant attenuation law for kiloparsec-sized star forming regions. Future work will include using the 
spectral information to test models of galaxies where star formation is being or has recently been extinguished.

We discuss the sample selection for the \swift+MaNGA (SwiM) catalog and its basic properties in Section~\ref{sec:sample}, and detail the \textit{Swift}/UVOT and MaNGA data reduction in Section~\ref{sec:sw_man_reduce}. The integrated photometric measurements are described in~\ref{sec:int_phot}. The process of spatial matching between Swift and SDSS images and MaNGA IFU spectroscopy is presented in Section~\ref{sec:sp_match}. We discuss the AGN fraction of the sample in Section~\ref{sec:agn} and provide notes about individual objects in Section~\ref{ssec:obj_notes}. We summarize in Section~\ref{sec3:conclusion}. We assume a $\Lambda$CDM cosmology when quoting masses, distances, and luminosities, with $\Omega_{\textrm{m}}=0.3$, $\Lambda = 0.7$ and $\textrm{H}_{0} = 70$~km~s\textsuperscript{$-1$}~Mpc\textsuperscript{$-1$}.
\section{The S\MakeLowercase{wi}M Catalog}\label{sec:sample}
\subsection{SDSS-IV/MaNGA and \textit{Swift}/UVOT}

Our sample of galaxies is a subset of the MaNGA survey \citep{Bundy2015,Yan2016}, which is a spatially resolved optical IFU spectroscopic survey included in the fourth generation of SDSS \citep[SDSS-IV;][]{Blanton2017}. The main sample of the MaNGA survey will have $\sim10,000$ galaxies that (1) create a uniform distribution in stellar mass for ${\rm M}_* > 10^9~{\rm M}_\odot$, as approximated via the SDSS $i$-band absolute magnitude, (2) provide uniform spatial coverage in units of half-light radius ($R_e$), and (3) maximize the spatial resolution and signal-to-noise for each galaxy \citep{Wake2017}. \redtext{This work uses the MaNGA Product Launch 7 (MPL-7) version of the MaNGA sample, which includes all objects observed by June 2017, totaling 4706 galaxies.}

The MaNGA data were taken with hexagonal IFU fiber bundles that
contain between 19 and 127 $2^{\prime\prime}$ fibers, extending over a diameter of between $12^{\prime\prime}$ and $32^{\prime\prime}$ \citep{Drory2015}. The fiber bundles are inserted into pre-drilled holes on plates mounted on the Sloan 2.5-m telescope \citep{Gunn2006}, and \redtext{fed} into the dual-channel Baryon Oscillation Spectroscopic Survey (BOSS) spectrographs \citep{Smee2013}. The BOSS spectrograph is designed with red and blue arms in order to provide continuous wavelength coverage from the NUV to near-IR\null. Therefore the MaNGA spectra span the wavelength range $3,622$--$10,350$~\AA\ with a resolving power of $R\sim2000$, allowing for the use of most nebular diagnostics, including BPT diagrams \citep{Baldwin1981}. The telescope is dithered to achieve near-critical sampling of the point spread function (PSF), which has been measured at $2\farcs 5$, while the resulting data cubes have a spatial sampling of $0\farcs 5$.  The total exposure time for each object is set by the sum of the squared signal-to-noise ratio: the sum of $(S/N)^2$ must be at least 20~pixel$^{-1}$~fiber$^{-1}$ in the $g$-band at galactic-extinction-corrected $g = 22$~AB mag, and 36~pixel$^{-1}$~fiber$^{-1}$ in the $i$-band at galactic-extinction-corrected $i= 21$~AB mag. These criteria result in an average integration time of 2.5 hours \citep{Yan2016}. 

In order to probe both NUV and optical properties, we also make use of \textit{Swift}/UVOT photometry. UVOT is a 30-cm telescope with a field of view (FoV) of $17^{\prime}\times17^{\prime}$, an effective plate scale of $1^{\prime\prime}$ pixel$^{-1}$, and three NUV filters: uvw2, uvm2 and uvw1 \citep{Roming2005}. While the UVOT PSF varies with wavelength (see Table~\ref{table:sw_spec}), all three filters give a resolution around $2\farcs 5$, i.e., similar to the angular resolution of the MaNGA optical spectra. The detector in the UVOT is a microchannel plate intensified CCD that operates in a photon counting mode, which can cause bright sources to suffer from coincidence loss \citep{Poole2008,Breeveld2010}. However, as discussed in Section~\ref{ssec:coincidence}, the galaxies in our sample are too faint for this to be a significant problem.\vspace{-2mm}
\begin{deluxetable}{lcccccc}[t!]
  \tablecaption{\textit{Swift}/UVOT NUV Observation Properties\label{table:sw_spec}}
\tabletypesize{\scriptsize}
\setlength{\tabcolsep}{1pt}
\tablehead{
\colhead{ } & \colhead{Central}& \colhead{Spectral} & \colhead{PSF} & \colhead{Median} & \colhead{Minimum} & \colhead{Faintest\vspace{-5pt}}\\
{ } &{Wavelength\tablenotemark{a,b}} & {FWHM\tablenotemark{a}} & {FWHM\tablenotemark{a}} & {Exposure} & {Exposure} & \colhead{Magnitude\tablenotemark{d}\vspace{-2pt}}\\
{Filter} & {(\AA)} & {(\AA)} & {(arcsec)} & {(s)} & {(s)} & {($m_{\textrm{AB}}$)}}
\startdata
{uvw2} & {1928} & {657} & {$2.92$} &{2375} & 187 & 22.3\\
{uvm2\tablenotemark{c}} &  {2246} & {498} & {$2.45$} &{2093} & {166} & 22.3\\
{uvw1} & {2600} & {693} & {$2.37$} &{1658} & 120 & 21.1\\
\enddata
\tablenotetext{a}{All filter properties are from \citet[][]{Breeveld2010}.\vspace{-2mm}}
\tablenotetext{b}{The central wavelength assumes a flat spectrum in $f_{\nu}$.\vspace{-2mm}}
\tablenotetext{c}{The uvm2 exposure time statistics only include galaxies with uvm2 images.}
\tablenotetext{d}{The faintest magnitude in our data set for each filter.}
\vspace{-1cm}
\end{deluxetable}

\subsection{The Cross-matched Catalog}
\label{sec:sample_prop}

We constructed our sample by cross-referencing the \redtext{MPL-7}, i.e., the SDSS Data Release 15 (DR15) \citep{Aguado2019}, with the \textit{Swift}/UVOT NUV archive as of April 2018. The UVOT archive is a combination of stars, active and normal galaxies, and gamma ray burst sources. The objects included in this sample are not always targeted, but instead fall within the FoV of UVOT. We required all objects to have usable, science-ready data cubes from MaNGA and have uvw1 and uvw2 observations in the \textit{Swift}/UVOT archive. We do not require uvm2 data for inclusion in the sample. Using these criteria, we obtained a sample of 150 galaxies, 87\% of which have uvm2 observations. We report the UVOT properties, and exposure time statistics, and the faintest magnitude detected in each filter in Table~\ref{table:sw_spec}.

The basic properties of the \swim\ galaxies, including the SDSS classification from DR15 \citep[see ][for details]{Bolton2012}, derived quantities, and observation information from MANGA and UVOT, are stored in a table available in its entirety in the electronic edition of the Astrophysical Journal. We show the format for this table in \redtext{Table~\ref{table:catalog_dm} in Appendix~\ref{app:vac_mod}, and the data model for the spatially-matched maps in Tables~\ref{table:maps_dm}--\ref{table:rawswft_dm} in Appendix~\ref{app:hdu_mod}.} 
57\% of the sample have either ``Galaxy'' or no available SDSS classification, while 31\% are classified as star-forming, 8\% as AGN and 4\% as starbursts. The median redshift is 0.033, which corresponds to a luminosity distance of 134.3~Mpc, and a spatial scale of 0.6~kpc$/^{\prime\prime}$. The stellar masses of the galaxies are provided by the NASA Sloan Atlas catalog \citep[NSA; ][]{Blanton2011} v1\_0\_1 based on the aperture-corrected elliptical-Petrosian photometry (for details see \citealt{Wake2017}). The stellar masses for this SwiM sample are in the range $8.73\leq\log(M_*/M_\odot)\leq11.11$, with a median of $\log(M_*/M_\odot)=10.02$. The SFRs are calculated by the MaNGA Data Analysis Pipeline \citep{Westfall2019} from the H$\alpha$ emission line flux within one effective radius, $1~R_e$, via the scaling relation given in \citet{Kennicutt2012}, after an internal reddening correction assuming the \citet{Odonnell1994} extinction law. The resulting SFRs span $0\lesssim\textrm{SFR(H}\alpha)\lesssim25~M_\odot~\textrm{yr}^{-1}$, with a median value of 0.18~$M_\odot$~yr$^{-1}$.

\subsection{Comparison of \swim\ to MaNGA}\label{ssec:sm_comp}
The MaNGA main sample is the combination of three different subsamples: Primary, Secondary, and Color-Enhanced samples \citep{Wake2017}. The catalog presented here contains 63 objects from the Primary sample, 57 from the Secondary sample, and 31 from the Color-Enhanced sample. Therefore, while these individually defined samples from MaNGA can be combined in different configurations, we compare the \swim\ to the combination of all three, i.e., the ``MaNGA main sample'' as defined in MPL-7. 

\begin{figure*}[t]
\centering
\includegraphics[width=0.48\textwidth]{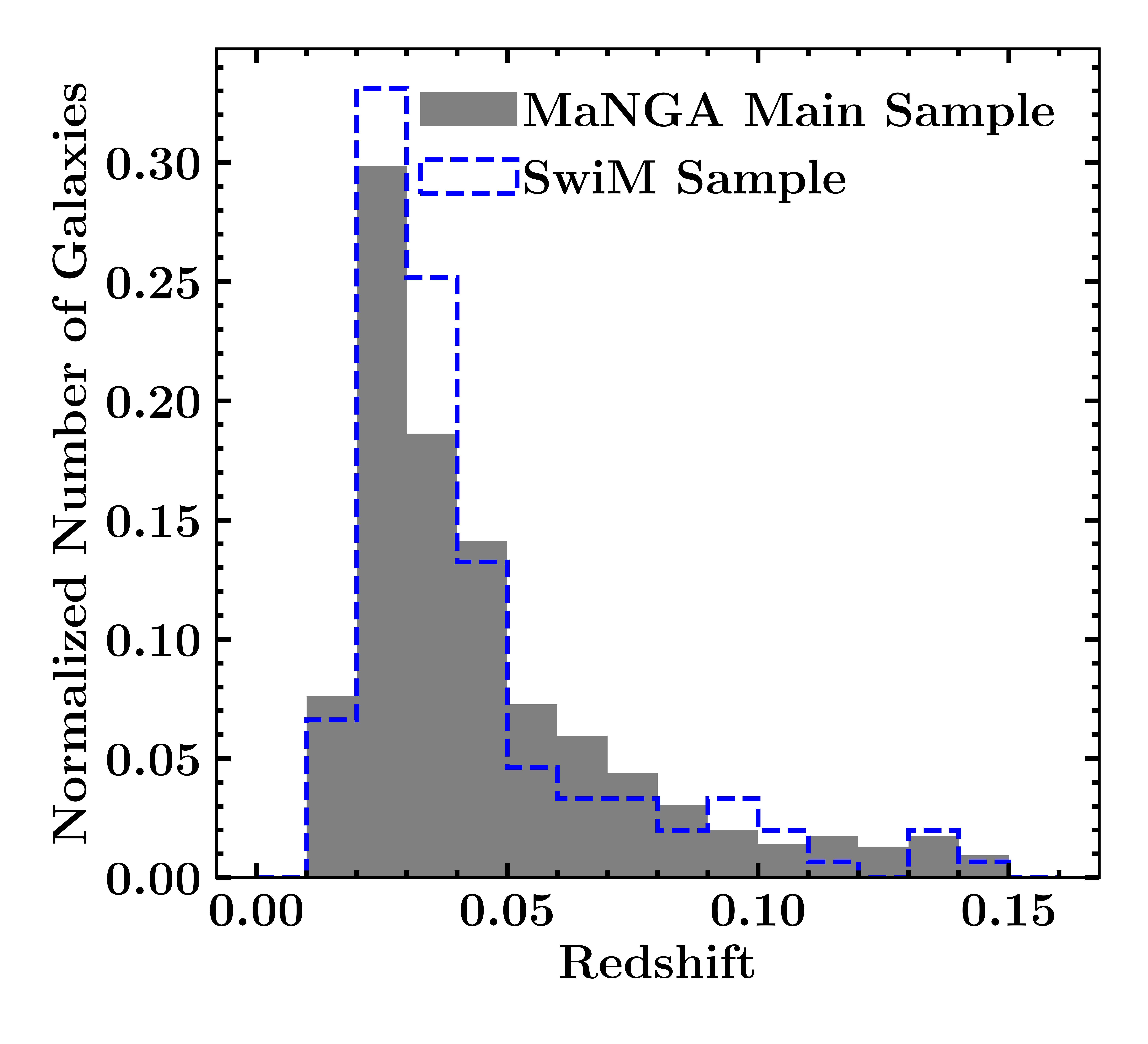}
\includegraphics[width=0.48\textwidth]{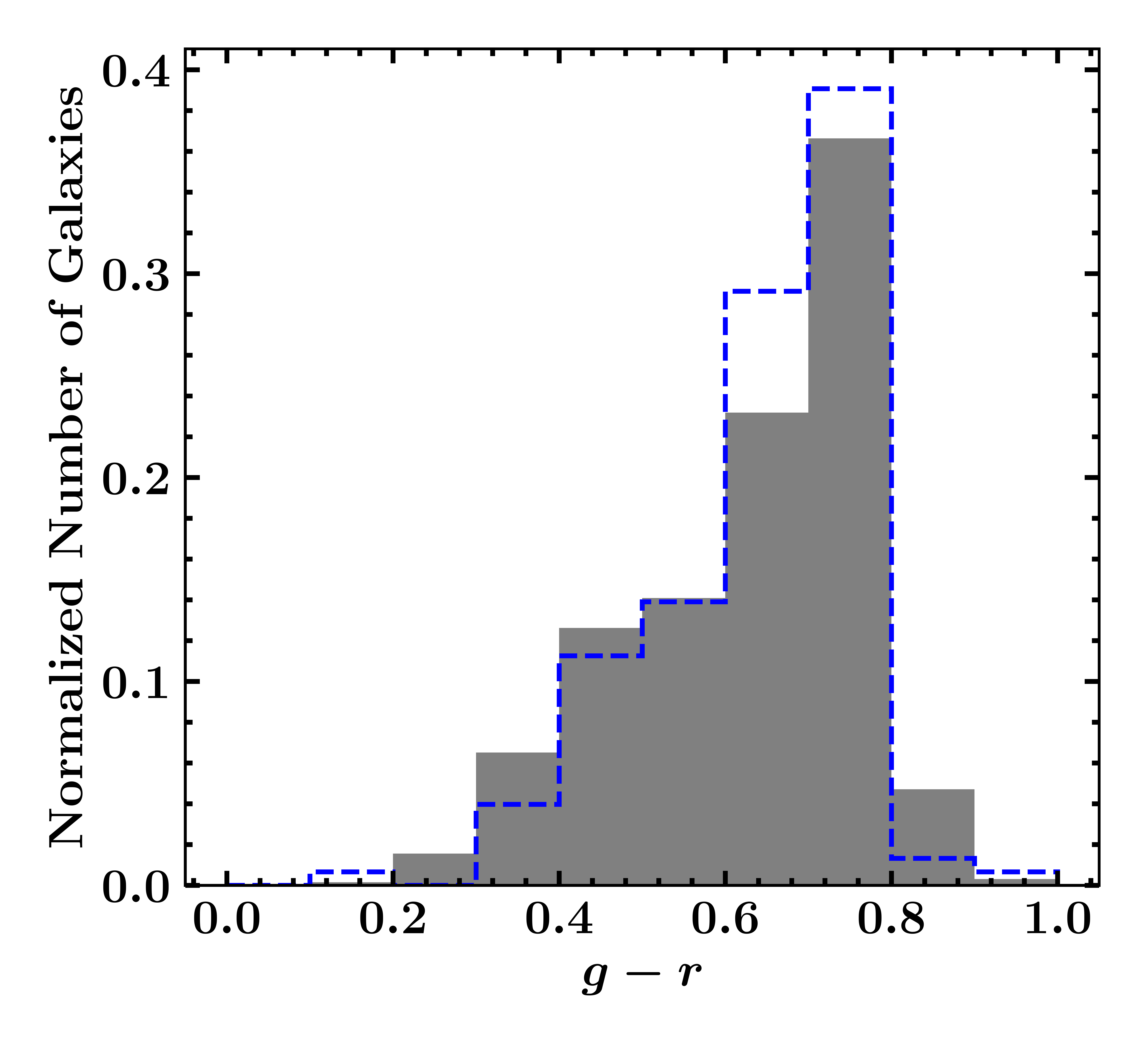}\vspace{-2mm}
\includegraphics[width=0.48\textwidth]{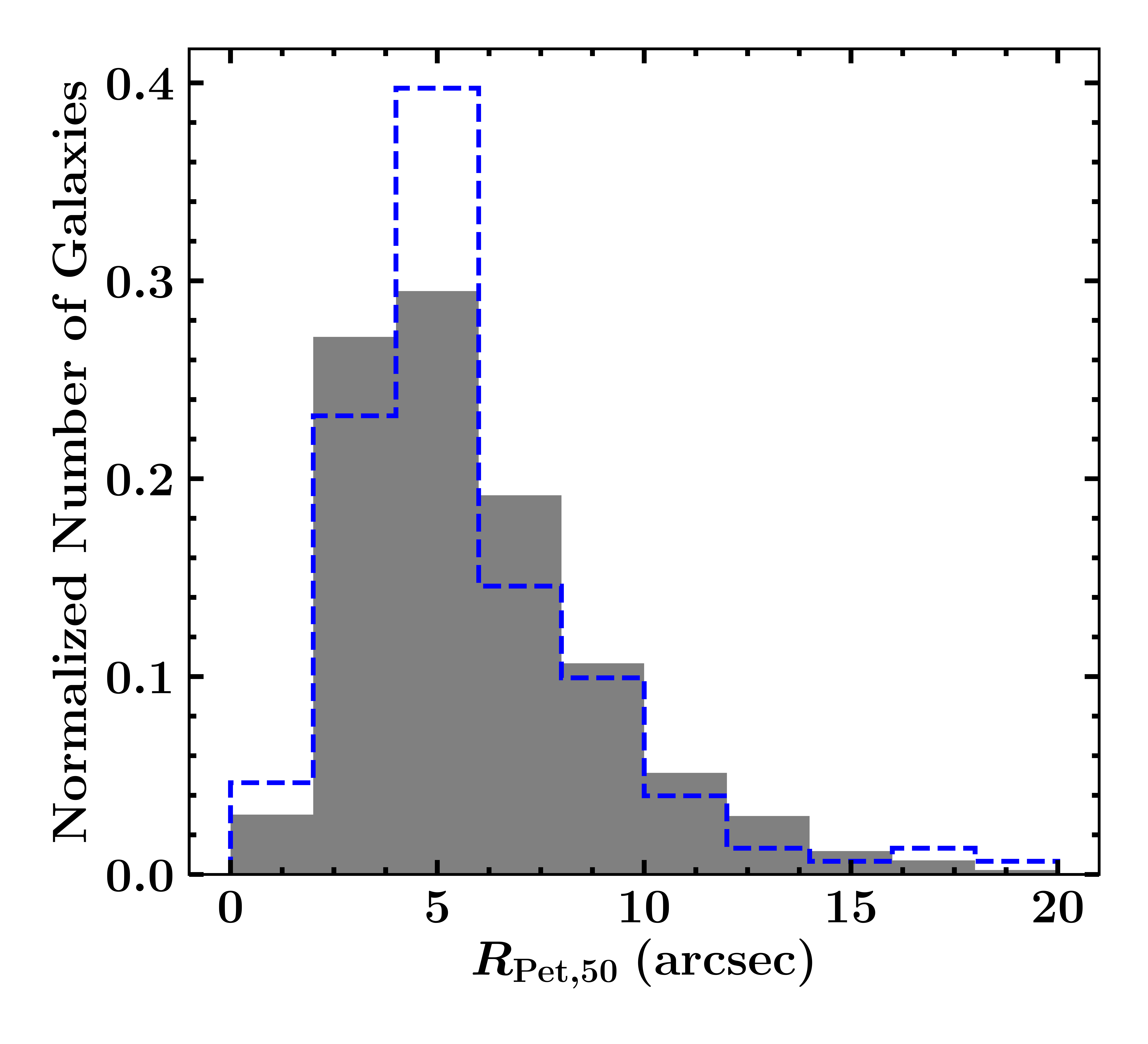}
\includegraphics[width=0.48\textwidth]{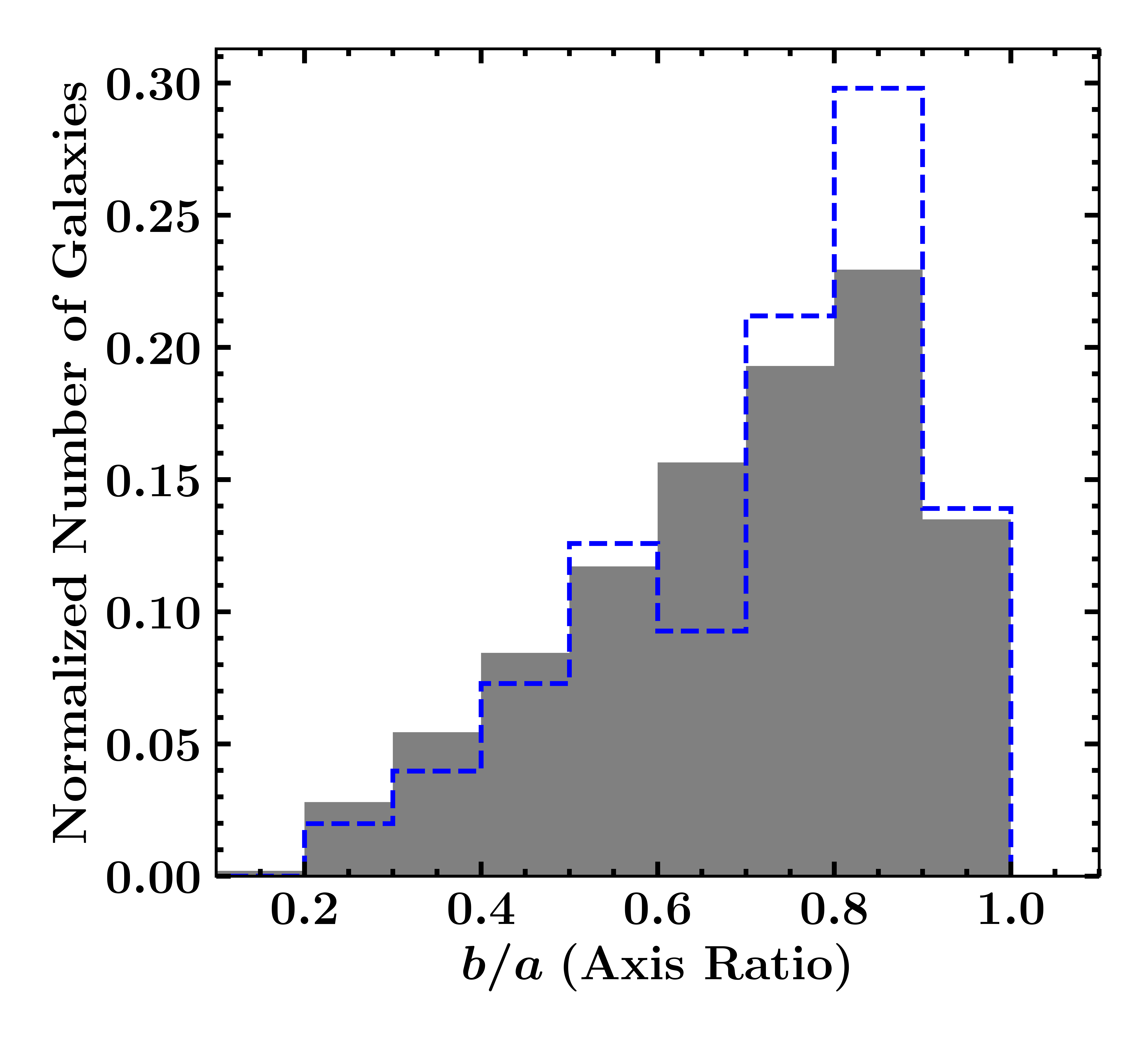}\vspace{-4mm}
\caption{Distribution of the redshift, \redtext{$g-r$ color}, Petrosian half-light radius (derived from $r$-band photometry), and the $r$-band axis ratio ($b/a$) of the galaxies in the \swim\ as compared to the MaNGA main sample. All histograms are normalized by the total number of galaxies in each data set. In each panel, the blue dashed outline represents the distribution of the \swim, while the gray shaded region represents the MaNGA main sample. The \swim\ has a similar distribution to the MaNGA main sample for all four properties. We perform a quantitative comparison of the two data sets in Section~\ref{sec:volwgt}. All data were taken from the NASA-Sloan Atlas.}\label{fig:histograms}
\end{figure*}

\begin{figure*}[t]
\hspace{-6mm} 
\includegraphics[width=0.48\textwidth]{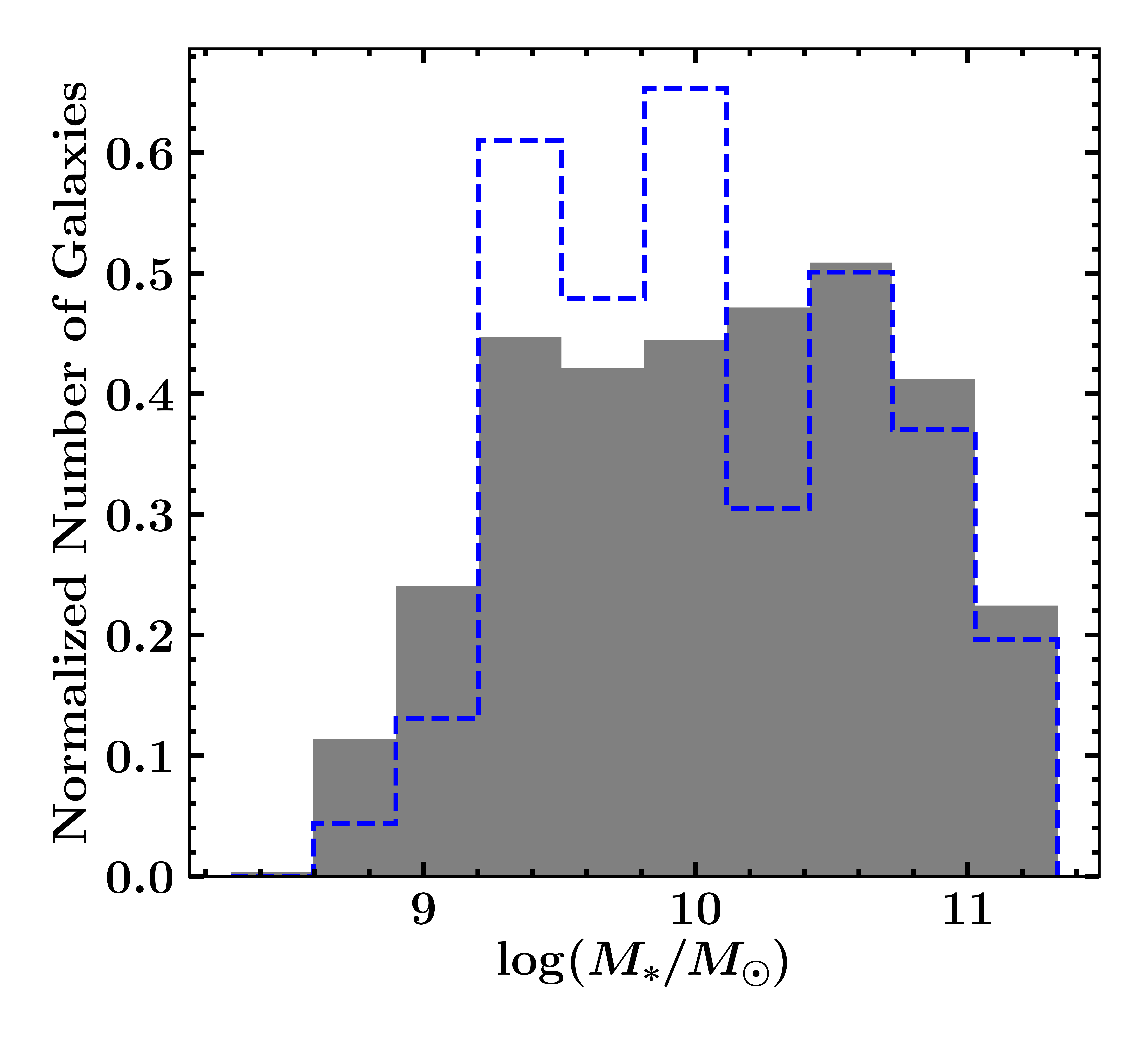}
\includegraphics[width=0.48\textwidth]{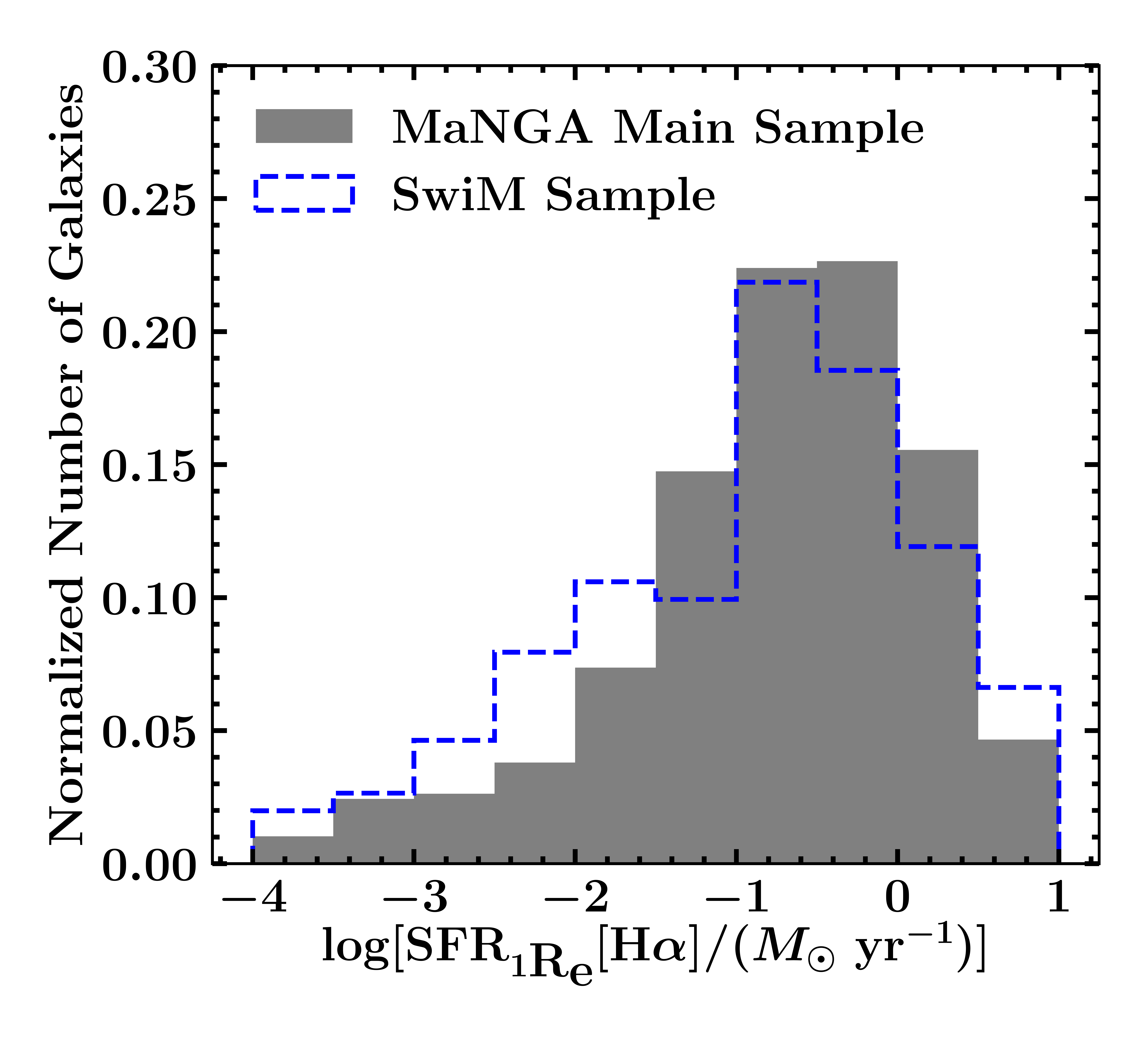}
\centering
\includegraphics[width=0.48\textwidth]{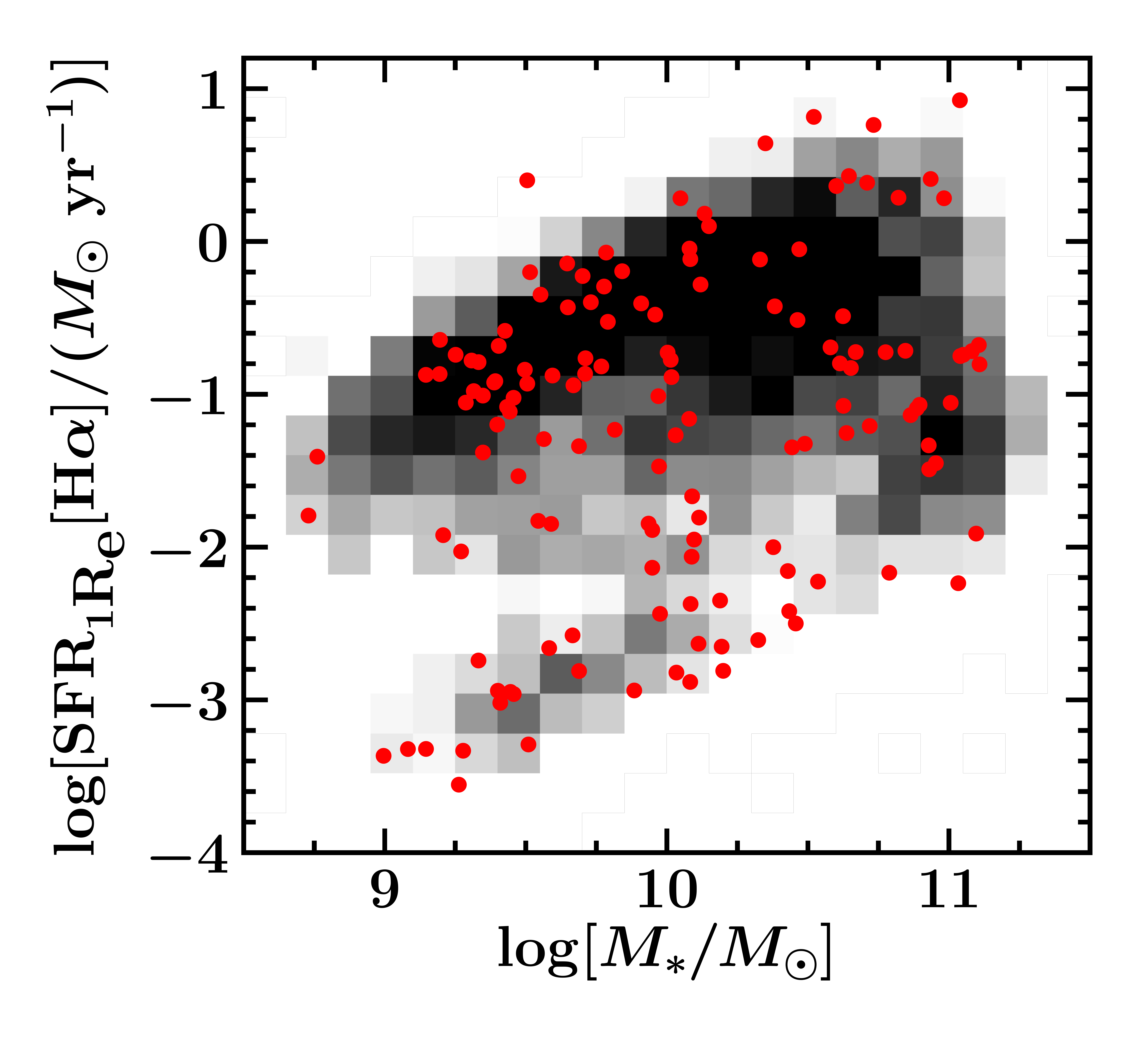}\vspace{-4mm}
\caption{\redtext{\textit{Top Left:} Same as Figure~\ref{fig:histograms}, but for stellar mass. The \swim\ has an over density of low-mass objects compared to the MaNGA main sample. \textit{Top Right:} Same as Figure~\ref{fig:histograms}, but for the SFR(H$\alpha$) within $1R_{\rm e}$. The $L$(H$\alpha$) measurements have been corrected for foreground extinction and internal attenuation in both catalogs, assuming the \citet{Odonnell1994} law and R$_V=3.1$. The \swim\ has an over density of low-SFR objects, and an under density of high-SFR objects compared to the MaNGA main sample. \textit{Bottom:}} SFR(H$\alpha$) within $1R_{\rm e}$ vs.~stellar mass for the \swim\ as compared to the full MaNGA sample. The $L$(H$\alpha$) measurements have been corrected for internal attenuation as described above. The catalog recovers the general distribution, except for the high-mass, high-SFR end of the star forming main sequence.}\label{fig:sfms_data}
\end{figure*}

We compare the distributions of redshift, \redtext{$g-r$ color}, size, and axial ratio of   \swim\ galaxies to those of the MaNGA main sample in Figure~\ref{fig:histograms}. The size is probed by the elliptical Petrosian half-light semi-major axis defined in the $r$-band, and all plotted quantities are taken from the NASA-Sloan Atlas. The majority of our galaxies fall within the range $z\lesssim0.05$, $R_{\rm Pet,50} < 20^{\prime\prime}$, and $b/a \gtrsim 0.6$. The distributions of redshift, $R_{\rm Pet,50}$ and $b/a$ in the \swim\ are similar to that of the MaNGA main sample. 
We also show the \redtext{distributions of stellar mass, the SFR and the} ``star-forming main sequence,'' i.e., the SFR (as given by H$\alpha$ within $1R_{\rm e}$) vs.~stellar mass for our catalog as compared to the full MaNGA sample in Figure~\ref{fig:sfms_data}. \redtext{The majority of our galaxies fall within the range of $\log(M_*/M_\odot)\gtrsim 9.5$.} The reduced MaNGA emission-line flux maps are already corrected for foreground Milky Way extinction, but not for internal attenuation in the target galaxy. While this can be complicated in the UV, most attenuation curves agree at redder wavelengths including H$\alpha$. We therefore assume the \citet{Odonnell1994} law, along with an intrinsic H$\alpha$/H$\beta$ ratio of 2.86 \citep[][chapter 11]{osterbrock2006}, and correct both the MaNGA main sample and the \swim\ for internal attenuation. \redtext{The corrected SFRs are shown in both the top right and bottom panels of Figure~\ref{fig:sfms_data}.} The \swim\ generally recovers the star forming main sequence seen in the MaNGA sample, \redtext{but is overly populated at the low-mass, low-SFR end, and} sparsely populated relative to MaNGA at the high-mass, high-SFR end. \redtext{This trend is also qualitatively seen in the SFR 1--D histogram on the top right of Figure~\ref{fig:sfms_data}.} 

\subsection{Correction Factors to Volume-Limited Weights}\label{sec:volwgt}

\begin{figure}[t]
\hspace{-8mm}
\includegraphics[width=0.5\textwidth]{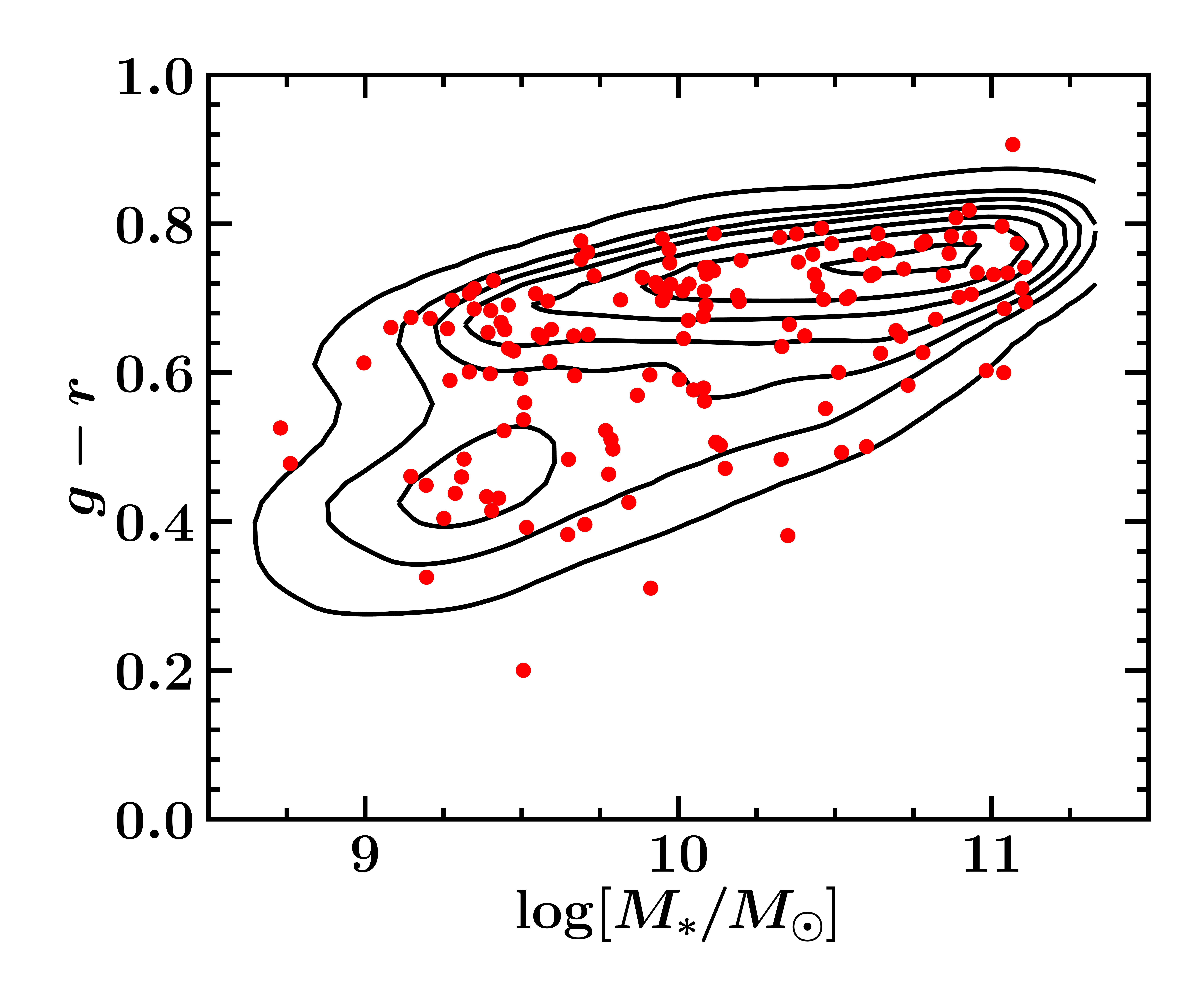}
\caption{The $g-r$ color vs.~stellar mass distribution of the entire MaNGA main sample, shown as black contours, compared to the SwiM sample, represented by red filled circles. Both quantities are measured within the elliptical Petrosian radius apertures and are from the NASA-Sloan Atlas \citep{Blanton2011}. The \swim\ sample captures the general distribution of the total MaNGA sample over our mass range of $8.73\leq\log({\rm M}_*/{\rm M}_\odot)\leq11.11$.}\label{fig:sfms_sample}
\end{figure}

\begin{figure}[t]
\hspace{-8mm}
\includegraphics[width=0.5\textwidth]{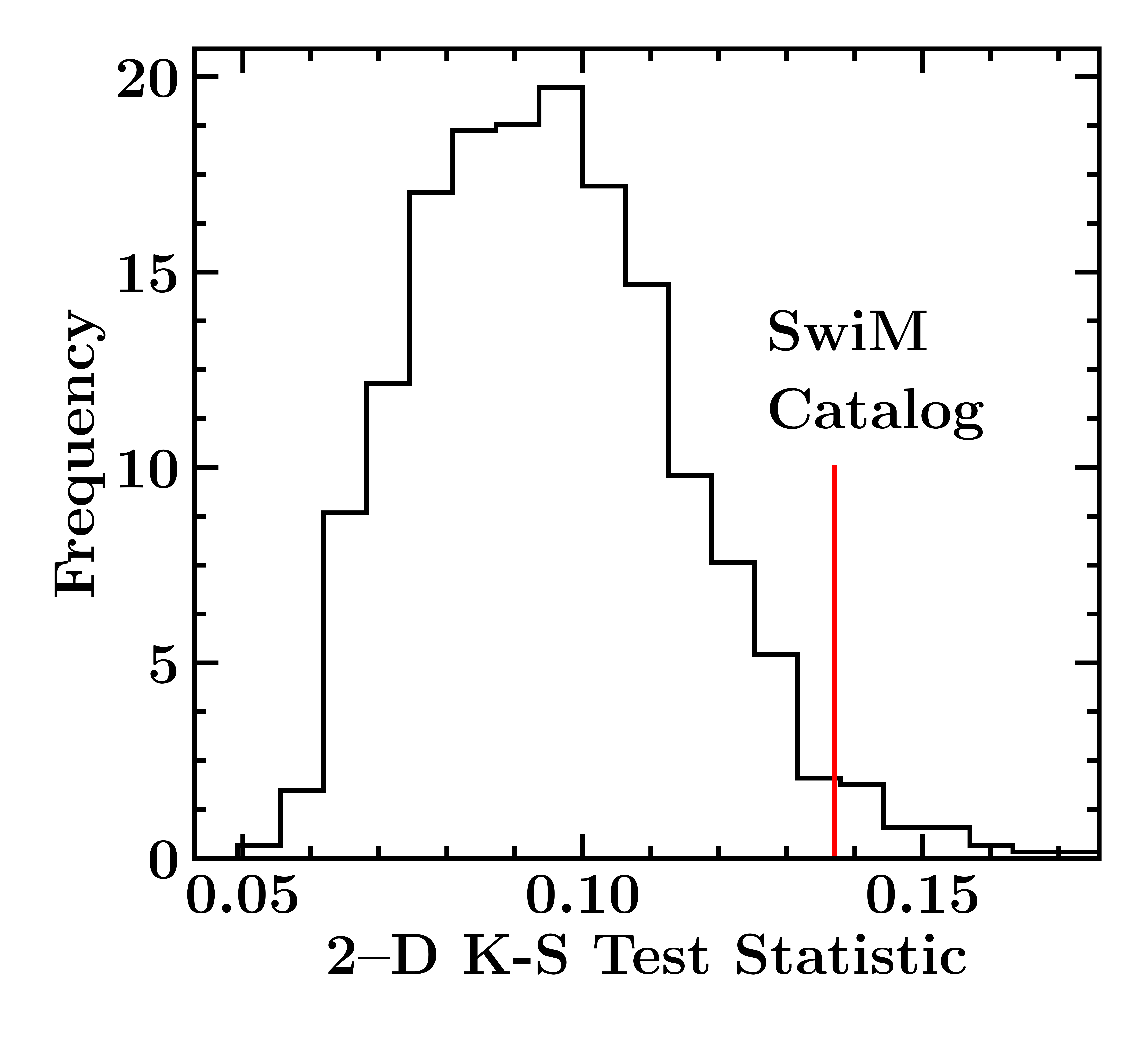}
\caption{Histogram of the 2--D K-S test statistic \citep{Peacock1983} for 1000 simulated random samples drawn from the MaNGA main sample. The test statistic for the \swim\ is denoted by the red vertical line. The \swim\ lies at the 96.1 percentile of the distribution, and therefore there is $3.9$ percent chance that a random sample pulled from the MaNGA main sample will have properties similar to the catalog.}\label{fig:rand_samp}
\end{figure}

The MaNGA sample was selected to have a flat number density distribution with respect to the stellar mass (as approximated by the SDSS $i$-band magnitude) and thus cannot be described as magnitude- or volume-limited. \citet{Wake2017} have provided weights for each MaNGA target that will statistically correct the sample to that of a volume-limited data set. If our catalog is consistent with a random sampling of the MaNGA main sample, then the weights calculated by \citet{Wake2017} should be applicable. We show the $g-r$ vs.~stellar mass distribution of the entire MaNGA sample as black contours, with the \swim\ overlaid as red points in Figure~~\ref{fig:sfms_sample}. While we sample a large portion of the parameter space, we do not have an even sampling of the MaNGA catalog in the color-mass space. We test for a quantitative similarity between our catalog and a random sample of the same size pulled from the MaNGA distribution using the 2--D K-S test \citep{Peacock1983}. Specifically, we create 1000 samples of 150 galaxies randomly drawn from the MaNGA main sample and compare the resulting 2--D K-S test statistic between that sample and the MaNGA main sample. We show the distribution of the test statistic, along with the measured test statistic for the \swim\ in Figure~\ref{fig:rand_samp}; the test statistic is defined such that a larger value denotes a lower probability that the two distributions are quantitatively similar in the observed 2--D parameter space. We find that the \swim\ lies at the 96.1 percentile of the distribution presented in Figure~\ref{fig:rand_samp}, and therefore there is only a $\sim4$\% chance that a random sample pulled from the MaNGA main sample would have properties similar to the \swim. We therefore conclude that there are some selection effects that make the \swim\ different from a randomly-drawn sample and the \citet{Wake2017} weights must be scaled in order to statistically correct our catalog to that of a volume-limited sample.

\begin{figure*}[h!]
\centering
\includegraphics[width=0.48\textwidth]{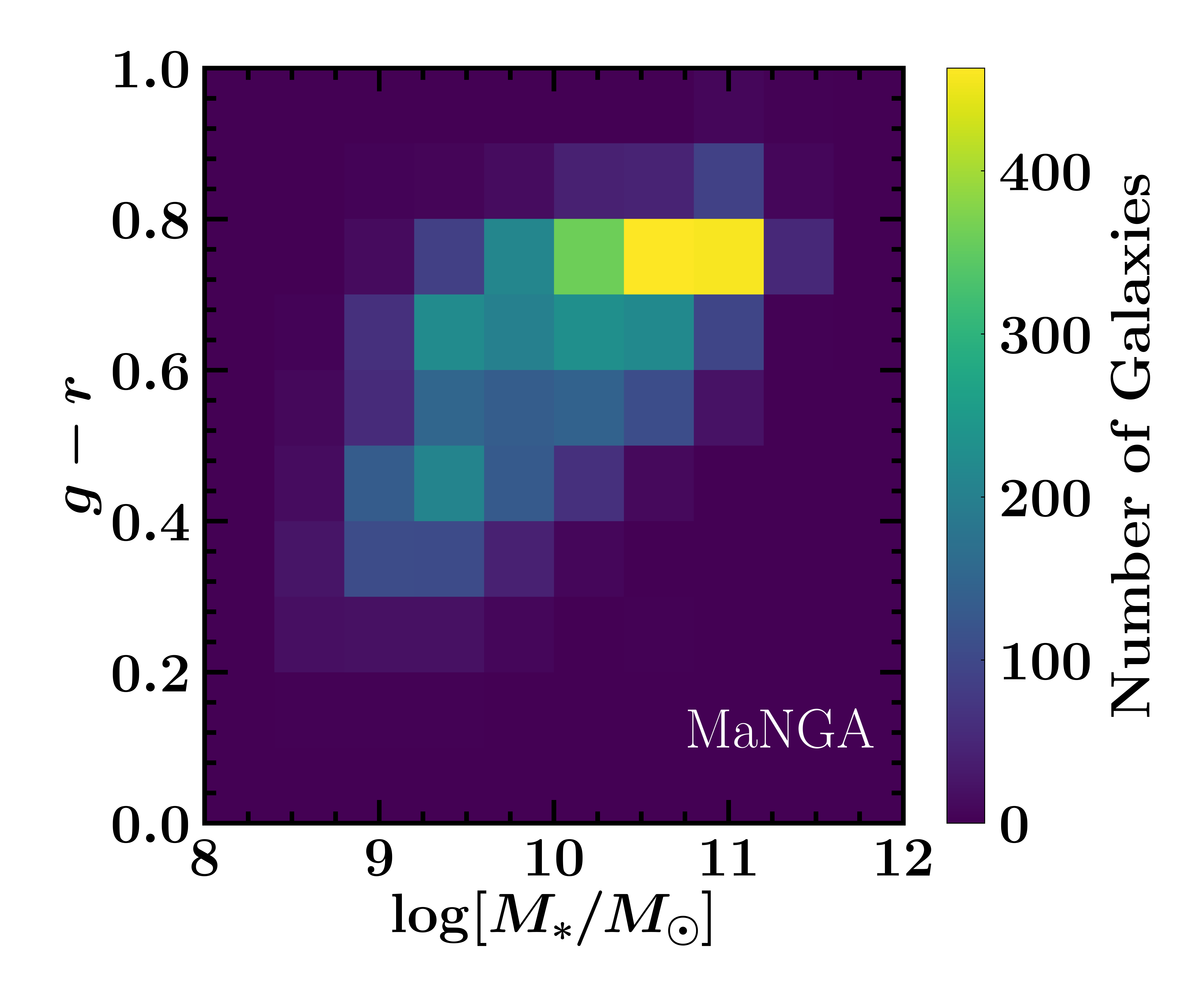}
\includegraphics[width=0.48\textwidth]{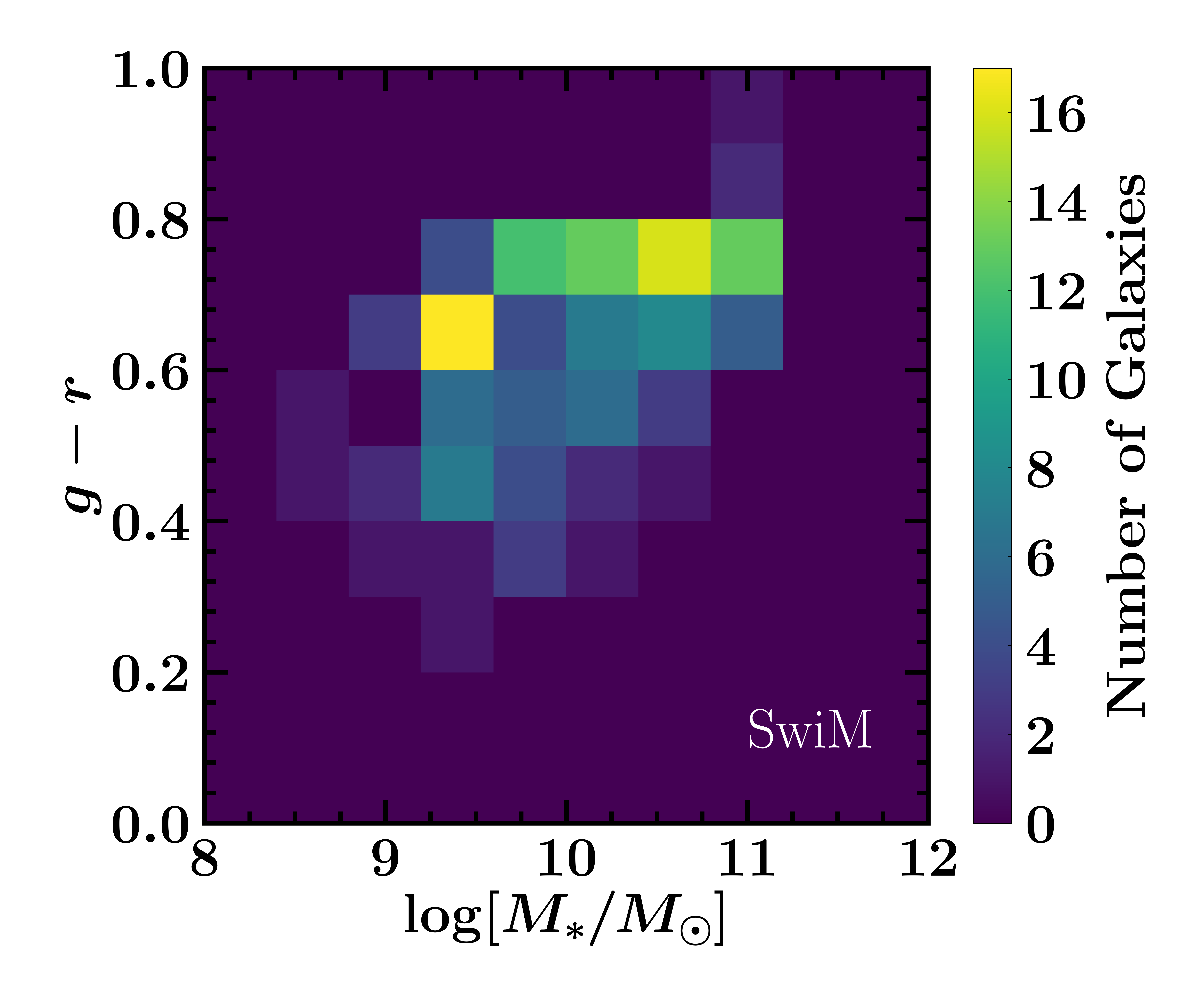}
\includegraphics[width=0.48\textwidth]{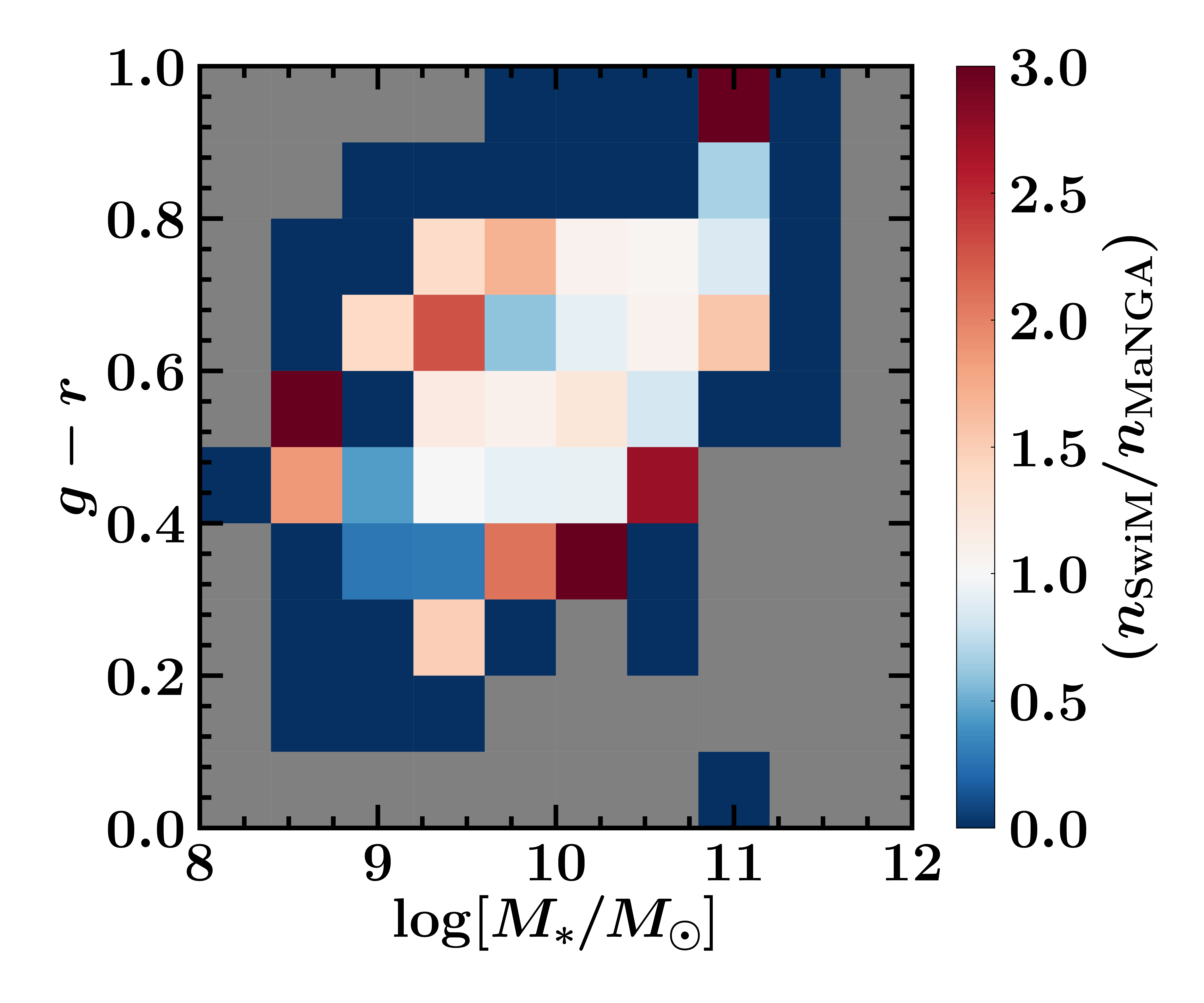}
\caption{2--D histograms of the number distribution of the MPL-7 MaNGA main sample (top left) and the \swim\ (top right), and the ratio between the two (bottom center), in $g-r$ vs.~stellar mass. The two sample distributions on the top show high density as yellow, and low density as shades of purple, denoted by the color bar. The MaNGA main sample has a strong peak in the high-mass portion of the red sequence and a secondary peak in the low-mass, blue portion of the diagram. Meanwhile the \swim\ samples the red sequence in a different way and includes a smaller fraction of low-mass blue galaxies. The bottom shows the ratio of the number densities of the \swim\ ($n_{\rm{SwiM}}$) and the MaNGA main sample ($n_{\rm{MaNGA}}$). Both number densities are calculated by normalizing to the total number of objects in each sample (150 and 4498, respectively). If the number densities are equal, the bin color is white, while over-densities in the \swim\ are represented by shades of red and under-densities by shades of blue, as denoted by the color bar. The qualitative number density differences between the two catalogs seen in the top two panels are quantified here.}\label{fig:2d_ratio}
\end{figure*}

In order to quantify those scaling corrections, we binned both the \swim\ and the MaNGA main sample using a linearly spaced, $10\times10$ binning scheme in the $g-r$ vs.~stellar mass space as shown in Figure~\ref{fig:2d_ratio}.  When plotted this way, the difference between the two data sets become obvious:  the MaNGA main sample has a strong peak in the high-mass red end of the diagram with a second weaker peak at the low-mass, blue end of the diagram. Meanwhile the \swim\ samples the red sequence in a different way (i.e., the band of higher density bins with $0.6\lesssim g-r\lesssim 0.8$), and lacks significant coverage of the low-mass blue end.  If we then divide the \swim\ and MaNGA main sample 2--D histograms, both normalized to the number of objects in the respective samples, we create the density ratio plot seen in the bottom panel of Figure~\ref{fig:2d_ratio}. \redtext{The differences between the two samples are quantified here:} the \swim\ is under-dense in the low-mass blue portion of the diagram and has regions of over- and under-density in the red sequence.

\redtext{The \swim\ is a subset of MaNGA and has a significantly lower number of total galaxies (150) than the MPL-7 MaNGA main sample ($\sim4700$). Therefore,} the distribution of the \redtext{un-normalized} fraction of MaNGA galaxies in the \swim\ in each bin (i.e., $N_{\rm{SwiM}}/N_{\rm{MaNGA}}$) is well-described by a binomial distribution. We use the un-normalized, binned 2--D distribution (the ratio of the two top panels in Figure~\ref{fig:2d_ratio}) to provide scaling factors in addition to the \citet{Wake2017} `esweights' or the volume weights for the MaNGA main sample. We present these ratios as the scale factors for each galaxy, along with the uncertainty as derived from the binomial distribution. To calculate the new weights, one should simply multiply the inverse of this scaling factor by the `esweight', both of which are provided in the Swim\_all catalog file. We do note that there are a significant number of bins, particularly on the edges of the 2--D distribution, that the \swim\ does not cover. In fact, 8\% of galaxies in the MaNGA main sample fall in bins that have no \swim\ galaxies. Thus, while these scaling factors can be used, there is some uncertainty in the correction that is highly dependent on the galaxy population of interest. We therefore caution users to only use scaling factors where the galaxies of interest lie within the bins populated by the \swim.

\section{\textit{Swift}/UVOT and MaNGA Data Reduction}
\label{sec:sw_man_reduce}
\subsection{\textit{Swift}/UVOT Pipeline}
\label{ssec:sw_pipeline}
All of the UVOT data in our catalog are archival \redtext{and} are drawn from observations obtained from the High Energy Astrophysics Science Archive Research Center (HEASARC). The UVOT data are processed using the \textit{Swift} UVOT Pipeline\footnote{\url{github.com/malmolina/Swift-UVOT-Pipeline}}, an automated and updated version of the subroutine \texttt{uvot\_deep.py} from the UVOT Mosaic program, written by Lea Hagen\footnote{\url{github.com/lea-hagen/uvot-mosaic}}. 

The \texttt{uvot\_deep.py} subroutine reads in data already downloaded from HEASARC, and follows the basic data processing procedures for UVOT images as described in the UVOT Software Guide\footnote{\url{heasarc.gsfc.nasa.gov/docs/swift/analysis}} as described below. The program ensures that both the counts and exposure maps are aspect corrected, reducing the uncertainty in the defined world coordinate system to $0\farcs 5$. Occasionally \texttt{uvot\_deep.py} will produce errors that cause the exposure map to have 0 or NaN values for small regions of pixels. This error only occurs in 5 galaxies out of the 150 galaxy sample. Furthermore, the error is only present within the galaxy itself for two objects. We provide masks to correct for this issue, which is described in Appendix~\ref{app:vac_mod}.

All UVOT images are mosaics of single frames with very short exposures that are stacked together to produce a deep image. UVOT does allow for different frame exposure times according to the science goal of the observation. However, the UVOT software will not combine frames with different frame times, as this would greatly complicate the analysis. Currently \texttt{uvot\_deep.py} requires the standard full frame exposure time of 11.0322~ms for inclusion in the final image. Additionally, all individual frames must be $2\times2$ binned, yielding a plate scale of $1^{\prime\prime}$ pixel$^{-1}$. If an individual frame meets both of these requirements and is aspect corrected, then it is added to the final image. As both criteria are standard for non-event mode data, we retain the majority of frames in this process. The final counts and exposure maps are corrected for large scale structure \citep[see][for details]{Breeveld2010} and have all bad pixels masked.

The \textit{Swift} UVOT Pipeline automates \texttt{uvot\_deep.py}, and will reduce multiple Swift images in a single execution. In one run, the pipeline will parse data already downloaded from HEASARC and run a modified version of \texttt{uvot\_deep.py}. This modified version will align the large scale structure correction map (if needed), skip any image that does not meet the requirements of the original \texttt{uvot\_deep.py}, and store that information in a log file for reference. This process is then repeated for each image of interest.

\begin{figure}[t]
    \centering
    \includegraphics[width=0.49\textwidth]{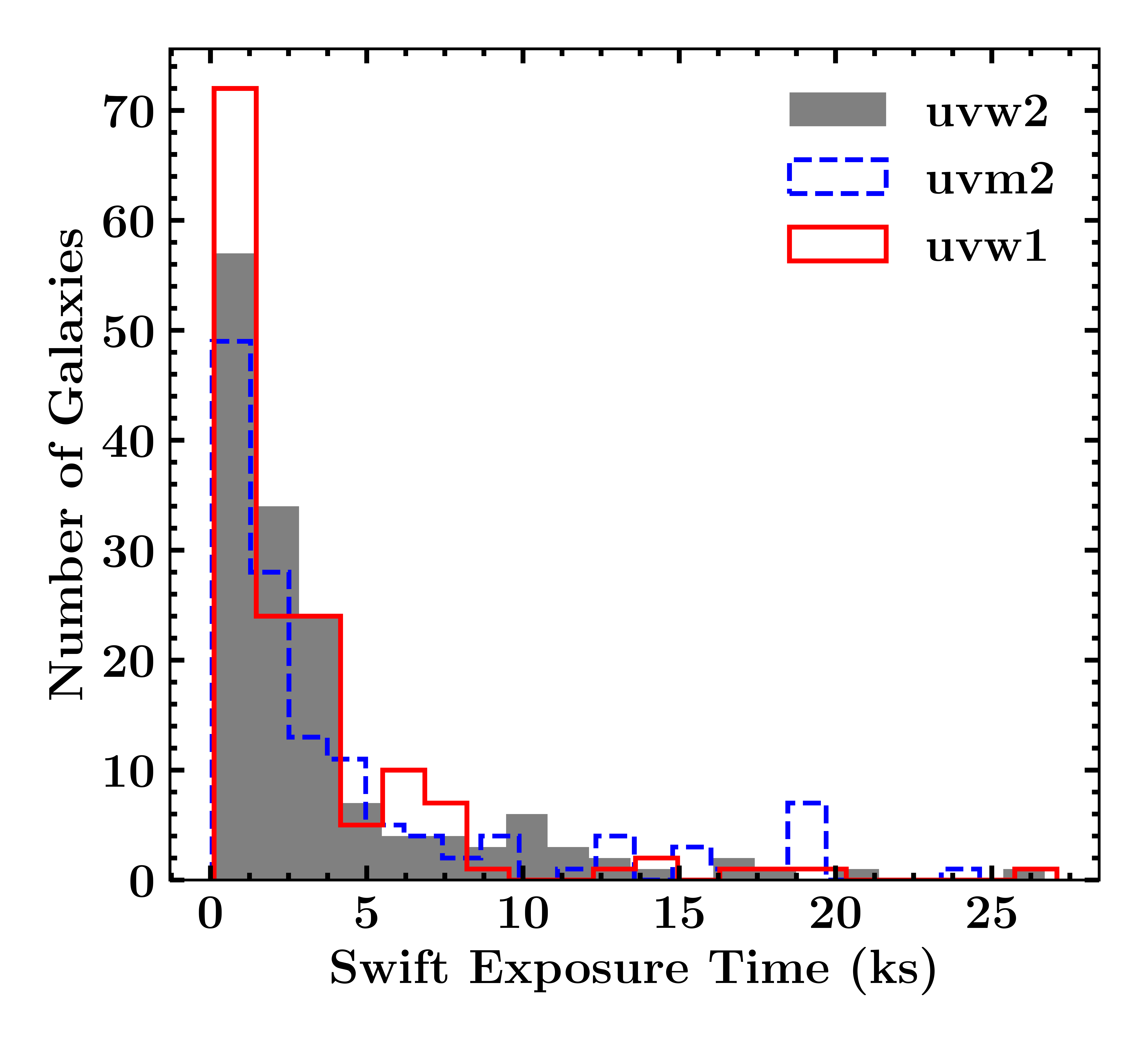}\vspace{-6mm}
    \caption{\redtext{The distribution of exposure times for each Swift/UVOT filter in kiloseconds. The exposure time distribution for uvw2 is shown in as the grey filled histogram, while that of uvm2 and uvw1 are shown in dashed blue and solid red lines, respectively. The majority of exposures in all three filters are less than 5~ks. The minimum visual magnitudes for a 5$\sigma$ detection are listed in Section~\ref{ssec:sw_pipeline}.}}
    \label{fig:sw_exp_hist}
\end{figure}

\redtext{The distribution of exposure times for the \swift/UVOT NUV observations are shown in Figure~\ref{fig:sw_exp_hist}. The majority of objects have exposure times of less than 5000~s in all three filters, with the median exposure time listed in Table~\ref{table:sw_spec}. The limiting visual magnitude for a 5$\sigma$ detection for a 5~ks observation is $m_{\rm V}=22.45$ for uvw2, $m_{\rm V}=22$ for uvm2, and $m_{\rm V}=22.2$ for uvw1. Additionally, the limiting visual magnitude for each filter, given their respective median exposures, are $m_{\rm V}=22.04$ for uvw2, $m_{\rm V}=21.53$ for uvm2, and $m_{\rm V}=21.6$ for uvw1.}

\subsection{Further UVOT Data Processing}

After the images are processed through the \textit{Swift} UVOT Pipeline, they are corrected for both the dead time and degradation of the detector. The UVOT detector is a microchannel plate intensified CCD, operating in a photon counting mode. As a result, approximately 2\% of the full frame time is dedicated to transferring charge out of the detector. This is corrected by increasing the count rate \citep{Poole2008}. Meanwhile, the decline in count rate due to the degradation of the detector is well characterized and provided in the UVOT calibration documents\footnote{\url{heasarc.gsfc.nasa.gov/docs/heasarc/caldb/swift/docs/uvot}}; this results in a 2.5\% correction for the most recent observations. 

Cosmic ray corrections are not necessary for UVOT images, due to its operation mode. For each $\sim11$~ms frame, all individual events are identified, and the centroid of the event location is saved. When the final image is created, each event is recorded as a count, with its location on the image given by the calculated centroid described above. In this regime, a cosmic ray that hits the detector will register at most a few counts in a single location, while a stationary astrophysical source will register thousands of counts. Therefore cosmic rays are incorporated into the background counts as they affect very few frames. 

\subsection{Coincidence Loss in \textit{Swift}/UVOT Images}\label{ssec:coincidence}
Coincidence loss occurs when two or more photons arrive at a similar location within the same $\sim11$~ms frame, causing a pile-up that affects the measured count rate. As UVOT operates in a photon-counting mode, this becomes a significant issue for bright sources. \citet[][]{Poole2008} characterized this effect for a single-pixel detector, while \citet[][]{Breeveld2010} describe coincidence loss when a point source is in front of a diffuse background (e.g., knots of star formation on top of a galaxy). Thus the \citet[][]{Breeveld2010} model appears to be the most appropriate for our data set.

However, in order to avoid PSF variation in individual filters, \citet[][]{Breeveld2010} recommended a minimum aperture size of $3^{\prime\prime}$ for all the three NUV filters, corresponding to a physical size of $\sim1.8$~kpc for the median redshift of our sample. With such an aperture, we cannot resolve individual \ion{H}{2} regions which have typical sizes of no more than a few hundred parsecs \citep[][]{Kennicutt1984,Garay1999,Kim2001,hunt2009}. Therefore, the point source plus diffuse background model described by \citet[][]{Breeveld2010} is not appropriate. 

Instead we approach this problem in the spirit of \citet[][]{Poole2008}. The formulated coincidence loss corrections are only valid for point sources, but the effect is insignificant when the count rate is below 10~counts~s\textsuperscript{$-1$}~pixel$^{-1}$. Across all of the UVOT observations in our sample the maximum count rate is $1.7$ counts s\textsuperscript{$-1$} pixel\textsuperscript{$-1$}.  This translates to a correction of $<0.2$\%, which is significantly smaller than the dead time correction of 2\%. Therefore the effects of coincidence loss are not significant for any of our UVOT observations and we ignore them in our catalog.

\subsection{\textit{Swift}/UVOT Sky Subtraction}

In order to quantify the local background in each UVOT image, we construct an annulus using two apertures, the inner elliptical aperture with a semi-major axis of twice the elliptical Petrosian semi-major axis \citep[$R_p$, from NASA-Sloan Atlas;][]{Blanton2011}, and the outer circular aperture with a radius of $4 R_p$. We measure the background counts with this annulus.

\redtext{We calculate the sky background using a three step process to better describe the local background emission. First, we mask all astrophysical contaminants within the annulus. Second, we mask all pixels within the annulus that do not have the same exposure time as the galaxy. After this processing, we use the biweight estimator \citep{Beers1990} on the remaining pixels within the annulus to calculate the final background counts for the galaxy. We describe each step in detail below.}

Given the size of the sky annuli, neighboring bright stars or galaxies may fall within it. To mitigate this effect, we run Source Extractor \citep{Bertin1996} and set the contrast parameter (DEBLEND MINCONT) to 0, which \redtext{identifies} 
even the faintest local peaks inside the annular region \redtext{and generates a background mask free of these objects.}

As \swift/UVOT images are made up of of individual frames that are stacked, there may be an uneven exposure map around the galaxy of interest. Thus, we only consider sky pixels with exposure times equal to that of the center of the galaxy. \redtext{We therefore mask out pixels that do not meet this criterion. This step ensures that all pixels used for the background calculation have the same observational properties as the galaxy, giving a more faithful measurement.}

\redtext{After applying both of these masks to the sky annulus, we measure the background counts on the remaining pixels} within the annulus using the biweight estimator \citep{Beers1990}. \redtext{Even with these masks, we retain a sufficient number of sky pixels for each galaxy to have a robust background measurement. Our masked annuli still have a median sky coverage of $\sim3300$--$3500$~square arcseconds, and a minimum value of $\sim800$~square arcseconds.}


\subsection{MaNGA}

The MaNGA spectra come fully reduced via the MaNGA Data Reduction Pipeline \citep[DRP;][]{Law2016}, which completes both the basic extraction and calibration steps needed to produce datacubes. These datacubes are then processed using the MaNGA Data Analysis Pipeline \citep[DAP;][]{Westfall2019}, which produces the best-fitting model spectra for all pixels that were successfully fit. The DAP also creates the 2--D maps of the measured emission line strengths and spectral indices, and measured quantities such as the H$\alpha$ emission within 1 effective radius. All MaNGA maps are corrected for foreground extinction using the $E(B-V)$ value from \citet{Schlegel1998}, and assuming the \citet{Odonnell1994} Milky Way dust extinction curve with $R_V=3.1$. 

We utilize the MPL-7 reduction version for the DRP and DAP, which is identical to that of SDSS DR15. The DAP utilizes a list of 42 average stellar continuum templates that was constructed by hierarchically-clustering templates from the MILES stellar library \citep{Sanchez2006}. For more details on this process, see Section~5 of \citet{Westfall2019}.
The emission lines and stellar continua in the MaNGA data cubes are fitted using the Penalized Pixel-Fitting code \citep[pPXF;][]{Cappellari2004}, and can employ two binning schemes. We use the ``HYB10'' scheme which is optimized for emission line measurements. This binning method involves first Voronoi-binning the data to calculate stellar kinematics, and then deconstructing the bins and performing emission-line and spectral-index measurements on the individual spaxels.

To increase both the efficiency and the accuracy of the templates, the DAP first combines all spectra in a given datacube into a single spectrum. That spectrum is fitted with a stellar continuum model using pPXF and the 42 templates described above. All templates that have non-zero weights are then used to conduct a second fit, this time to the Voronoi-binned data \citep{Cappellari2003}. This fit provides the stellar kinematic information, of which the first two moments are saved. The finalized template is then stored and used in the subsequent fitting of the individual spaxels described below.

The emission lines and stellar continua in each spaxel are fitted using a two-step process: first the emission lines and single finalized template from the previous step are fitted to the individual spaxels, assuming a Gaussian profile with identical velocities and velocity dispersions. This step determines the optimal stellar-continuum template and adjusts the initial guesses for the emission-line kinematic components. Finally, the spectra for the individual spaxels are fitted again, this time allowing the velocity and velocity dispersion of the individual emission lines to vary, while adopting known physical constraints for all doublets. This creates the final emission-line model. The spectral indices are measured from the individual spectra after subtracting the best-fitting emission-line model. However, we recalculate the Lick indices and D$_n$(4000) measurements to allow for binned measurements, as described in Section~\ref{ssec:lickidx}.   


\section{Integrated Photometric Measurements}\label{sec:int_phot}

We present the observed AB apparent magnitudes in the \galex\ FUV,  \textit{Swift}/UVOT NUV and SDSS optical filters for all the galaxies in the \swim\ in the SwiM\_all catalog file. The \galex\ and SDSS measurements come from the NASA-Sloan Atlas \citep{Blanton2011}, while those from \swift/UVOT are measured using our dataset. We measure the integrated photometry for these bands using the same aperture used by NSA v1\_0\_1, which is the $r$-band elliptical Petrosian aperture. 
These integrated magnitudes need to be corrected for the light lost outside the aperture due to the instrument PSF\null. This could be a larger effect in \galex\ and \swift/UVOT, as their PSFs are wider than the SDSS images.  
The NASA-Sloan Atlas already corrects the \galex\ and SDSS photometry for this issue. Thus, we must apply a similar correction to the UVOT NUV measurements. The corrections are completed in a three-step process: first the galaxy's $r$-band 2--D light profile, as projected on the sky, is modeled to create a simulated galaxy. The model is then convolved with the PSF of the UVOT NUV filter of interest (uvw2, uvm2, or uvw1) to simulate an observation in that filter. Finally the original $r$-band integrated measurement from SDSS is compared with that of the simulated galaxy that was ``observed'' by UVOT in order to calculate the fraction of light lost. This procedure is completed for each galaxy and the corrections are applied to the elliptical $r$-band Petrosian integrated galaxy UVOT NUV magnitudes.

The photometric measurements are \textit{not} corrected for either foreground extinction or internal attenuation, and are not K-corrected. 
\section{Spatial Matching of SDSS Data Products to \textit{Swift}/UVOT}\label{sec:sp_match}

In order to enable a joint analysis using SDSS imaging, Swift/UVOT imaging, and a MaNGA spectral datacube, we have to transform all the images and maps to the same spatial resolution and spatial sampling. The Swift/UVOT uvw2 filter has the coarsest PSF ($2\farcs 92$ FWHM), as shown in Table~\ref{table:sw_spec}. Thus, we need to convolve all other images and maps to this resolution. To match the spatial sampling, it is inevitable that noise and covariance will be introduced during the resampling process.  To minimize this effect, we choose to keep the data with the lowest S/N intact and apply the resampling to the data with the highest S/N. The Swift/UVOT data generally have lower S/N compared to SDSS imaging and MaNGA spectra, and have the coarsest sampling with 1\arcsec\ pixels. Moreover, the uvw2 photometry tends to have a lower S/N than uvw1 photometry. Thus, we choose to resample all data to match the sampling in the uvw2 band. We describe this process below for all of the quantities of interest. 

Given the relatively low S/N of the UV data, further binning is likely necessary to make use of this dataset. We thus strive to present the final data in a format that would facilitate binning of the end user's choice. 

\subsection{SDSS and \textit{Swift}/UVOT Images}

The uvw2 images are kept in the original format. We convolve each of the uvw1, uvm2, and SDSS $u,g, r,i,z$ images with an appropriate kernel to match the PSF in the uvw2, and then reproject them to the uvw2 sampling. We set the convolution kernel to a 2D Gaussian with 
\begin{equation}\label{eqn:kernelsize}
\sigma= \sqrt{({\rm FHWM}_{\rm uvw2} - \epsilon)^2- {\rm FWHM}_{\rm x}^2 \over 8\ln 2}  , 
\end{equation} 
where FWHM$_{\rm x}$ represents the FWHM of the PSF of the corresponding filter, and $\epsilon$ is a correction term we will discuss below. In general, PSFs are not Gaussians; they are closer to Moffat functions, which have additional power-law wings.  However, given that most of our sources are faint, only the core and not the additional wings are strongly detected. We therefore approximate the PSFs as Gaussians. The FWHM for uvw1 and uvm2 are given in Table~\ref{table:sw_spec}. For SDSS images, we use a FWHM of $1 \farcs 4$, typical of the seeing condition for the SDSS data. Small variations in the seeing will not significantly change the kernel size. 

We then reproject the convolved images for uvw1, uvm2, and SDSS $u,g,r,i,z$-bands to the pixel positions in the uvw2 image using the flux-conserving spherical polygon intersection algorithm. This is achieved by using the {\it reproject.reproject\_exact} function in {\it astropy} \citep{Whelan2018}. 

This reprojection process brings an additional broadening to the effective PSF. With simulations, we found that this additional PSF broadening varies depending on the amount of shift in the pixel grid. For uvw1 and uvm2 with 1\arcsec\ pixels, the broadening can vary from 0 to 0.1\arcsec\ in the $\sigma$ of the PSF for a Gaussian with a FWHM of 2.92\arcsec. A polynomial fit as a function of fractional pixel shifts could predict this broadening effect to better than 0.001\arcsec. Thus, for each galaxy, for uvw1 and uvm2 filters, we apply the corresponding correction factor ($\epsilon$) in Equation ~\ref{eqn:kernelsize} depending on the amount of fractional pixel shift between the pixel grids. For MaNGA and SDSS, due to their smaller pixels, the PSF-broadening effect is smaller and shows much less variation with fractional pixel shifts. For SDSS, the broadening varies from 0.03\arcsec\ to 0.04\arcsec\ with a median around 0.0376\arcsec\ for $\sigma$. For MaNGA, the broadening varies from 0.03\arcsec\ to 0.05\arcsec\ with a median around 0.0419\arcsec\ for $\sigma$. In both of these cases, we use the median correction for all galaxies. The amount of remaining error is at most 0.012\arcsec\ in $\sigma$ and 0.028\arcsec\ in FWHM. This is less than 1\% of the final PSF width and is negligible, as the measurement error of the PSF is usually larger than this.

The exposure maps and masks are also processed in the same way through the convolution and reprojection. Masked pixels are ignored in the computation.  
The final processed mask is rounded to 0 or 1 using a threshold of 0.4 . If more than 40\% of a pixel area comes from bad pixels, then the final pixel is considered bad (mask=1). 

\subsection{Spatial Covariance for Swift/UVOT and SDSS images}\label{sec:sdss_covar}
The convolution and reprojection introduce covariance between neighboring pixels. This not only means the final uncertainty is larger than that computed based on error propagation without including covariance, but it also means that the final images contain covariance. If one were to bin the final images further, one would need to take this covariance into account when estimating the photometric errors.

We compute the final covariance matrix in the following way. 
For SDSS and Swift uvw1 and uvm2 images, we start by constructing a covariance matrix with only diagonal elements containing the variance of all the pixels, basically assuming all pixels are independent of each other. We refer to this matrix as $G$.  Then we construct the matrices corresponding to the convolution and reprojection processes, which we refer to as $W$ and $Z$, respectively. For an image with an initial size $N\times N$ and a final reprojected size $M\times M$. The $W$ matrix has the shape of $N^2 \times N^2$, and the $Z$ matrix has the shape of $M^2 {\rm (rows)} \times N^2 {\rm (columns)}$ . The final covariance matrix can then be computed as 
\begin{equation}\label{eqn:covar_propagation}
    C = (Z\times W) \times G \times (Z\times W)^T 
\end{equation}

The final covariance matrix has a size of $M^2 \times M^2$. The diagonal elements of $C$ contain the variance for the final images. We then take the square root to obtain the 1-$\sigma$ uncertainty map. 

\begin{figure}
    \centering
    \includegraphics[width=0.5\textwidth]{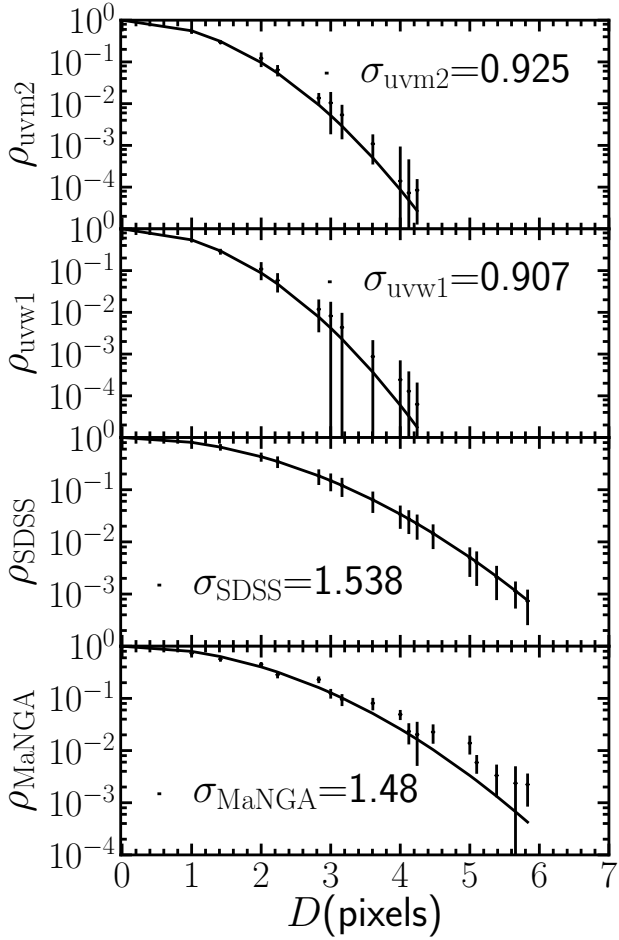}

    \caption{The spatial correlation strength in the final maps as a function of distance between pixels, for galaxy 1-44745. The data points and error bars show the mean and standard deviation for each distance bins. The curves show the Gaussian fits over the plotted range (with amplitude fixed to 1 and centered at 0). The four panels, from top to bottom, show the results for uvm2, uvw1, SDSS images, and MaNGA maps, respectively. The correlation can be well described by Gaussian functions with the scale indicated in the legend.}
    \label{fig:correlation}
\end{figure}

The final map still contains covariance. This can be characterized by the correlation length. We recast the covariance matrix, $C$, to the correlation matrix, $\rho$, by computing $\rho_{ij}= C_{ij}/\sqrt{C_{ii}C_{jj}}$ for all $i$ and $j$  from 1 to $M^2$. The correlation matrix has all diagonal elements equal to 1. Other elements give the correlation strength between each pair of pixels. Only pairs of pixels with small spatial distance have non-zero values. Following the example given by \cite{Westfall2019} and using galaxy, MANGID 1-44745, as a typical example, we find the correlation can be well fit by a Gaussian function of the pair-wise distance. The scale parameter of the Gaussian is 0.925 for uvm2, 0.907 for uvw1, and 1.538 pixels for SDSS filters.  These are shown in Figure~\ref{fig:correlation}.   

\begin{figure}
    \centering
    \includegraphics[width=0.5\textwidth]{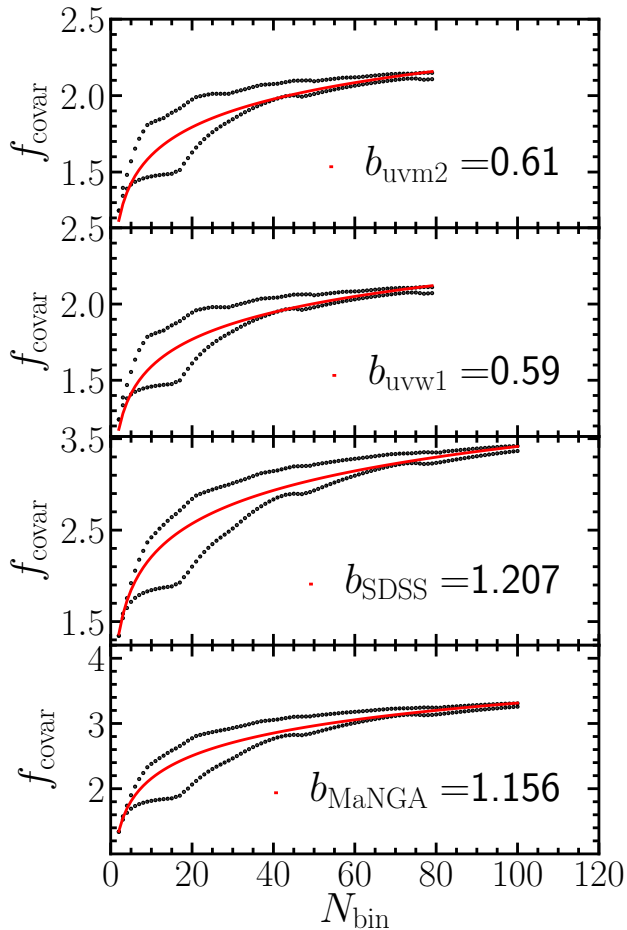}
    \caption{Scaling factor between errors propagated with and without including covariance, as a function of the number of pixels binned together. The black dots show the simulated difference between the two sets of errors for the two extreme cases of binning, as described in Section~\ref{sec:sdss_covar}. The red curves show the fit using the functional form of $f_{\rm covar} = 1+ b\log (N_{\rm bin})$. The four panels, from top to bottom, are for uvm2, uvw1, SDSS, and MaNGA maps, respectively.}
    \label{fig:fcovar}
\end{figure}

In lieu of providing the full covariance matrix, here we provide a functional form as an approximation for the effect of the covariance. We construct a mock map with unity errors in all pixels and bin $N$ pixels together and propagate the errors in two ways, with and without taking covariance into account. We plot their ratio, $f_{\rm covar} = \sigma_{\rm covar}/\sigma_{\rm no\_covar}$, as a function of the number of pixels binned together. Because the correlation is between neighboring pixels, the effect of the covariance depends on the shape of the bin. We compute two extreme cases to bracket different situations. The maximum covariance case is for a bin that is nearly a square, similar to the case of Voronoi binning. The minimum covariance case is for a long rectangular bin (with a maximum length of 27 pixels), similar to the case of annular binning. Figure ~\ref{fig:fcovar} shows how $f_{\rm covar}$ scales with the number of pixels in the bin, under these two cases. We fit them using the same functional form as suggested by \cite{Husemann2013}. The fit results are listed below. For large $N_{\rm bin}$, the scaling factor asymptotes to a constant value. \\
UVM2: 
\begin{equation}\label{eqn:m2_covar}
f_{\rm covar} = \left\{\begin{array}{cc}
         1+ 0.61\log (N_{\rm bin})& \mbox{if $N_{\rm bin}<80$} \\
         2.16 & \mbox{ otherwise}
    \end{array} \right.
\end{equation}
UVW1:
\begin{equation}\label{eqn:w1_covar2}
f_{\rm covar} = \left\{\begin{array}{cc}
         1+ 0.59\log (N_{\rm bin})& \mbox{if $N_{\rm bin}<80$} \\
         2.12 & \mbox{ otherwise}
    \end{array} \right.
\end{equation}
SDSS:
\begin{equation}\label{eqn:sdss_covar}
f_{\rm covar} = \left\{\begin{array}{cc}
         1+ 1.207\log (N_{\rm bin})& \mbox{if $N_{\rm bin}<100$} \\
         3.41 & \mbox{ otherwise}
    \end{array} \right.
\end{equation}

There is a caveat for the uncertainty maps of the SDSS images. Many inverse variance images in the NASA Sloan Atlas contain features due to satellite tracks. But these features do not appear in the flux images. If they are not masked in the inverse variance images, they would become the dominant feature in the final uncertainty maps. We applied additional masking to remove these features in our processing.

\subsection{MaNGA Emission Line Maps}\label{ssec:manmatch}
For emission line fluxes and equivalent widths (EWs), we start from the Gaussian-fitted 2-D emission line flux and EW maps generated by the MaNGA DAP \citep{Westfall2019}. Because the EW is a ratio between the line flux and the continuum, all the convolution and reprojection steps should be carried out on the line flux and continuum images first, before deriving the EW at the uvw2 resolution and sampling positions. 

By taking the ratio between the line flux and EW maps from DAP, we first derive the continuum map at the original MaNGA resolution and sampling positions. We convolve the flux and continuum maps with an appropriate Gaussian kernel to match the PSF of the \textit{Swift}/UVOT uvw2 filter. 
The maps were then reprojected to the \textit{Swift}/UVOT uvw2 pixel positions using the {\it reproject.reproject\_exact} function in astropy. 
Masked pixels are ignored in the computation and the final masks are produced by convolving and resampling the mask, which is then rounded to 0 or 1 to produce the final mask. 
If users of the catalog wish to bin the data further, the best approach is to divide the flux maps by the EW maps and then bin flux and continuum separately before computing the binned EW. 
We provide these measurements for all 22 emission lines provided by the DR15 version of MaNGA DAP \citep{Westfall2019}. 

\subsection{MaNGA Spectral-Index Maps}
\label{ssec:lickidx}
 
Spectral-index maps, including the Lick indices and D$_n$(4000) maps, are very useful for constraining the stellar populations of galaxies. Producing these properties at the uvw2 resolution requires more care as they all represent ratios of flux densities.
 
For example,  D$_n$(4000) is the ratio of the average flux density per unit frequency ($f_\nu$) between a red band (4000-4100\AA) and a blue band (3850-3950\AA\null). Simply presenting the final convolved D$_n$(4000) maps is not sufficient to allow further binning. Therefore, we have remeasured the blue and red band flux densities in the data using the DRP LOGCUBE files. We then convolved and resampled them to the Swift uvw2 PSF and pixel coordinates following the same process as done to the emission line flux maps. Instead of presenting their ratio in the final file, we provide the two flux density maps for each galaxy. 
The variance maps are also processed in the same way, and the final 1-sigma uncertainty maps are presented for each flux density map. The mask for D$_n$(4000) is derived from the DAP mask for D$_n$(4000), and is processed in the same way as that for emission line maps. 
 
In order to allow flexibility in further binning of the resulting map, we chose a different definition of the Lick indices from the standard definitions adopted by the MaNGA DAP \citep[see Section 10 of][]{Westfall2019}.  The standard definitions given by \cite{Trager1998},  define the continuum as a sloped line between the two side bands. It then integrates the fractional deficit in flux over the central band. Because the ratio between flux and continuum is inside the integral and the denominator is not a constant, this definition is inconvenient for spatial binning. Under this definition, proper spatial binning would require one to go back to the spectral datacube, bin the spectra, and then remeasure the indices.  
 
To allow more convenient spatial binning, we adopt an older definition of Lick indices, which is first used by \cite{Burstein1984} and described in detail by \cite{Faber1985}. Instead of a sloped continuum, this definition adopts a constant as the continuum in calculating the integral of the fractional flux deficit, as in 
\begin{equation}
    {I_{a} =} 
	   \begin{dcases}
	    	\int\left(1-\frac{f_\lambda}{f_{\rm C0}}\right)\,d\lambda & \mbox{in \AA} \\
	    	\\
		    -2.5\log_{10}\left(\frac{1}{\Delta\lambda}\int\frac{f_\lambda}{f_{\rm{ C0}}}\,d\lambda\right) & \mbox{in magnitudes}
	   \end{dcases}
		    \label{eqn:E4}
\end{equation}
Here, $f_{\lambda}$ is the total flux density per unit wavelength in the index band, $f_{\rm C0}$ is the continuum flux density per unit wavelength and $\Delta\lambda$ is the width of the index band. The value of $\Delta\lambda$ for all Lick indices is given in Table~\ref{table:specindx_dm} of Appendix A. Here $f_{\rm C0}$ is not a function of $\lambda$ and can be taken out of the integral. Therefore, Equation~\ref{eqn:E4} can be simplified to
\begin{equation}
     I_{a} =
	   \begin{dcases}
	    	{\Delta\lambda-\frac{F_I}{f_{\rm C0}}} & \mbox{in \AA} \\
	    	\\
		    -2.5\log_{10}\left(\frac{1}{\Delta\lambda}\frac{F_I}{f_{\rm{ C0}}}\right) & \mbox{in magnitudes}
	   \end{dcases}
		    \label{eqn:E4.1}
\end{equation}
 where $F_I$ is the integrated flux in the index band ($F_I=\int f_\lambda d\lambda$). With the continuum flux density taken out of the integral, one could get binned Lick indices without having to go back to the spectra, as long as both the continuum and the integrated flux in the passband is provided for each spaxel, and the constant continuum is defined to be strictly additive when spectra are added together. To get binned Lick indices, one would simply bin both the map of the continuum and the map of the flux before dividing them. 
 
 To define the continuum that is strictly additive, we first define the average flux density in the red and the blue bands as the following,
\begin{equation}
    f_{\rm R} = {1\over(\lambda_{\rm 2R} - \lambda_{\rm 1R})} \int_{\lambda_{\rm 1R}}^{\lambda_{\rm 2R}}{f_\lambda\, d\lambda}
    \label{eqn:E1}
\end{equation}
\begin{equation}
    f_{\rm B} = {1\over (\lambda_{\rm 2B} - \lambda_{\rm 1B})} \int_{\lambda_{\rm 1B}}^{\lambda_{\rm 2B}}{f_\lambda\, d\lambda}
    \label{eqn:E2}
\end{equation}
where $\lambda_{\rm 1R}$, $\lambda_{\rm 2R}$, $\lambda_{\rm 1B}$ and $\lambda_{\rm 2B}$ are the end points of the red and blue bands. We define the linear continuum flux as,
\begin{equation}
    f_{\rm C0} = (f_{\rm R} - f_{\rm B})\frac{\lambda_{\rm IM}-\lambda_{\rm BM}}{\lambda_{\rm RM}-\lambda_{\rm BM}} +f_{\rm B}
    \label{eqn:E3}
\end{equation}
where $\lambda_{\rm RM}$ and $\lambda_{\rm BM}$ are the mid-points of the red and the blue bands and $\lambda_{\rm IM}$ is the mid-point of the index band. This continuum level would be strictly additive when multiple spaxels are combined.

This definition of the Lick indices is also quite useful when it is applied to composite stellar population models. One simply has to measure the continuum and the integrated flux in the index band for each simple stellar population with a certain age and metallicity. When constructing the composite models, the Lick indices can be computed by adding the flux and the continuum separately before computing the index for the composite model. 

Under this definition, we need only to provide the maps for $F_I$, $f_{\rm C0}$, the associated uncertainty, and masks at the Swift UVOT resolution and sampling positions. For each spaxel, we take the spectrum from the MaNGA DRP LOGCUBE file, subtract from it the best-fit emission-line spectrum, then  transform it to the rest-frame given the redshift and the stellar velocity provided by DAP\null. We measure for each spaxel the index band integral and continuum for each Lick index, using the passbands of the MaNGA DAP \citep{Westfall2019}. We then convolve the resulting maps to the same PSF as the Swift uvw2, and reproject it to the Swift uvw2 pixel positions.   

\subsection{Spectral Resolution for the Lick indices}

The values of the Lick indices depend on the spectra resolution of the spectra. the stellar velocity dispersion of the target, and ``beam smearing'' resulting from any systematic variation in stellar velocities within the aperture used for the measurement. The traditional Lick-index system is defined for a constant instrumental resolution of 8.4~\AA\ FWHM, and a fixed stellar velocity dispersion. This instrumental resolution is too coarse for the higher resolution spectra from SDSS and MaNGA. We also argue that it is undesirable to smooth the data to match a fixed velocity dispersion, or to make an approximate and model-dependent correction using a fitting formula based on an object's velocity dispersion. To complicate the matter further, the instrumental resolution of the BOSS spectrograph varies with wavelength, whether it is specified in wavelength units or velocity units. A better approach is, therefore, to smooth the model spectra to match the combined effective dispersion in the data, which includes both the instrumental dispersion and the stellar velocity dispersion. Therefore, we provide, as part of our data products, the maps of the combined dispersion for each Lick index. 

The combined dispersion is constructed by adding in quadrature the stellar velocity dispersion with the instrumental dispersion for each spaxel and each index. The instrumental dispersion is taken at the center of the index band. For the convolution and reprojection process, we apply the square of the combined dispersion and weight the computation by the integrated flux in the index band. We also propagate the uncertainties and provide the associated masks. 

\subsection{Spatial Covariance in the MaNGA maps}\label{sec:man_covar}

The uncertainty maps for MaNGA emission line and spectral index properties are also produced by taking into account the covariance. In contrast to the Swift and SDSS images, the MaNGA maps come with significant covariance between spaxels. This means the $G$ matrix in Eqn.~\ref{eqn:covar_propagation} contains non-zero off-diagonal elements. To construct $G$ for MaNGA, we start by creating a correlation matrix ($\rho$) using a correlation scale of 1.92 spaxels, as provided by \cite{Westfall2019} (see Fig. 8 in that paper). We then multiply this matrix by the reformatted variance maps of each MaNGA property to build the covariance matrix, $G$, by $G_{\rm ij} = \rho_{\rm ij}G_{\rm ii}G_{\rm jj}$. The rest of the steps are similar to those described in Section \ref{sec:sdss_covar}.  

The final correlation in the resulting maps has a scale factor of 1.48 pixels, as shown in the last panel of Figure~\ref{fig:correlation}. This is for an example galaxy with MANGAID, 1-44745. For different galaxies, with different PSFs in the MaNGA data cube, the results could differ slightly. If one wants to bin the map further, to include covariance in the error propagation, one should use the $f_{\rm covar}$ given below to scale the error propagated without covariance. The fits for this covariance factor are shown in the last panel of Figure~\ref{fig:fcovar}
\begin{equation}\label{eqn:man_covar}
f_{\rm covar} = \left\{\begin{array}{cc}
         1+ 1.156\log (N_{\rm bin})& \mbox{if $N_{\rm bin}<100$} \\
         3.31 & \mbox{ otherwise}
    \end{array} \right.
\end{equation}

\subsection{Organization of the Maps}
Most of the per-galaxy data we provide are in the form of maps. Given that all the images and maps are convolved to the same PSF and reprojected to the same sampling positions, they also share the same World Coordinate System. We group these images and maps into several groups: broadband images, emission line fluxes, Lick indices, and D$_n$(4000). Each group contains images in multiple broadband filters, multiple emission lines, or multiple indices. We stack all the images and maps in each group together in a 3D array with different channels (layers) corresponding to different filters/lines/features.  The uncertainty and masks are also presented in corresponding 3D arrays.  All these arrays are presented as different header data units (HDU) in a FITS file with the extension name indicating the group and whether the file contains measurements, uncertainties, or a mask. The detailed data model is given in Appendix~\ref{app:vac_mod}. 

\subsection{Uncertainties of MaNGA-based measurements}\label{sec:manerr}

We provide formal errors associated with the data in this value-added catalog (VAC)\null. However, these formal errors could be underestimated or overestimated. It is much more reliable to use repeated observations to evaluate the uncertainty. Using repeated observations, the MaNGA team \citep{Belfiore2019} evaluated the uncertainty associated with the emission line flux measurements, and found the actual uncertainty is only slightly larger than the formal error, by 25\% for H$\alpha$, and by similar levels for other strong emission lines. Therefore, to get a realistic error estimates, one simply has to multiply the H$\alpha$ flux and EW errors by 1.25. \redtext{For more information on this process, please see \citet{Belfiore2019}.}

Similarly, for D$_n$(4000), \cite{Westfall2019} showed that a realistic error estimate based on repeated observation is about 1.4 times that of the formal error. Flux calibration systematics could be one of the contributing factors. However, here, we do not scale our error estimates for D$_n$(4000) because we are presenting the errors associated with the red band and the blue band separately. We recommend that the users propagate the formal errors to the final error for D$_n$(4000) and then multiply it by 1.4. 

For Lick indices, one could also derive these scaling factors. \cite{Westfall2019} found that the error scaling factor is 1.2 for H$\beta$ and H$\delta_A$ absorption EW, 1.6 for the Fe5335 index, 1.4 for the Mgb index, and 1.5 for the NaD index. 
  
\section{Active Galaxies in the S\MakeLowercase{wi}M Catalog}\label{sec:agn}
\subsection{\redtext{Identification of AGN}}
Because our only requirement for inclusion in the sample is the availability of \swift/UVOT data, there are AGNs present in the catalog. \redtext{We present our AGN identification method in this section, while providing notes on individual objects in Section~\ref{ssec:obj_notes}.} We searched for these AGNs using a three step process. We first used the spatially-resolved BPT diagrams from MaNGA to identify all objects with at least 10 MaNGA 0.5\arcsec-pixels within $0.3~R_e$ that fall within the Seyfert, LINER, or AGN regions of the [\ion{S}{2}]/H$\alpha$, [\ion{N}{2}]/H$\alpha$ or [\ion{O}{1}]/H$\alpha$ BPT diagrams. These are combined with all objects that have an SDSS classification of ``AGN'', ``QSO'' or ``Broadline'' to make the subset of 47 AGN candidates that are used in steps two and three.

Second, we identify all objects with detectable X-ray emission by utilizing archival data from the \textit{Swift} X-Ray Telescope \citep[XRT;][]{Burrows2000}, as all UVOT images have a corresponding XRT observation. The \textit{Swift}/XRT data come from the UK \swift\ Science Data Centre. The X-ray properties of all detected objects are obtained via the automated spectral fitting web tool \citep{Evans2009}, while the upper limits either come from the automated light curve web tool \citep{Evans2007,Evans2009} or the XRT point source catalog \citep[1SXPSC;][]{Evans2013}. We have 100\% coverage of our sample and 17 galaxies have detectable X-ray emission. While hard X-ray emission can be indicative of AGN, both low- and high-mass X-ray binaries (XRBs) can also produce hard X-rays. The PSF of \swift/XRT is $18^{\prime\prime}$ at 1.5~keV, which encompasses the entire galaxy for almost all of the objects in our sample. Thus, the XRBs present in the galaxy are contributing to the observed X-ray emission.  The contribution from low-mass XRBs is proportional to the stellar mass, while the contribution from high-mass XRBs is proportional to the SFR \citep{Fabbiano2006,Lehmer2010}. Therefore the observed X-ray emission must be stronger than the contribution from XRBs in order to be ascribed to an AGN. We calculate the XRB contribution using the stellar mass and SFRs for the galaxy presented in SwiM\_all catalog file, and the $L^{\rm gal}_{\rm HX}$ calculation given in equation (3) of \citet{Lehmer2010}. We report the photon index for the assumed power law, the unabsorbed 0.3--10~keV luminosities of the galaxies with detectable X-ray emission, and the XRB contribution from the galaxy in Table~\ref{table:xray}. While there is a well-defined relationship between the H$\alpha$ and the X-ray luminosities in AGN, e.g., \citet{Panessa2006}, the MaNGA maps do not include the broad H$\alpha$ component. In order to complete this test, detailed measurements of the broad components of AGN spectra must be made separately, which is beyond the scope of this paper. Due to short exposure time in the X-ray for the undetected objects, their upper limits are in the  range $L(0.3$--$10)\sim10^{41}$--$10^{43}$~erg~s$^{-1}$ and are not very meaningful. Thus, we do not report them in this paper. 

\begin{deluxetable*}{lccccc}
\footnotesize
  \tablecaption{X-Ray Detections in the \swim \label{table:xray}}
\tablehead{ {} & {} & \colhead{N$_{\rm H}$\tablenotemark{a}} & {$L_{\rm obs}$\tablenotemark{a}} & {$L^{\rm gal}_{\rm HX}$\tablenotemark{b}} & {Consistent}\\
\colhead{Object I.D.} & {$\Gamma$\tablenotemark{a}}& {(cm$^{-2}$)} & {($10^{40}$ erg~s$^{-1}$)}& {($10^{40}$ erg~s$^{-1}$)} &{with AGN?}}
\startdata
{1-37336} &  {$2^{+4}_{-1}$} & {$<5\times10^{21}$} & {$300^{+10000}_{-200}$} & {0.99} & {No\tablenotemark{c}}\\
{1-90242} &  {$1.5^{+0.3}_{-0.3}$} & {$<5\times10^{20}$} & {$3600\pm800$} & {...} & {Yes\tablenotemark{d}}\\
{1-95092} &  {$1.1^{+0.8}_{-0.5}$} & {$<5\times10^{21}$} & {$90^{+50}_{-30}$} & {0.53} & {Yes}\\
{1-109152} & {$-1.3^{+0.3}_{-0.2}$} & {$<7\times10^{20}$} & {$1800^{+300}_{-200}$} & {0.74} & {Yes}\\
{1-137883} & {$3\pm3$} & {$5^{+5}_{-4}\times10^{23}$} & {$4700^{+3\times10^5}_{-4300}$} & {0.14} & {Yes}\\
{1-153627} & {$2.1\pm0.4$} & {$<20\times10^{20}$} & {$24^{+9}_{-6}$} & {0.59} & {Yes}\\
{1-155975} & {1.7\tablenotemark{e}} & {$1.5\times10^{20}$\tablenotemark{e}} & {$21\pm8$} & {0.91} & {Yes}\\
{1-210784} & {$1.8\pm0.4$} & {$<20\times10^{20}$} & {$450^{+90}_{-80}$} & {0.22} & {No\tablenotemark{f}}\\
{1-269227} & {$2.0\pm0.1$} & {$10\pm1\times10^{21}$} & {$4400^{+500}_{-400}$} & {0.12} & {Yes}\\
{1-317315} & {$1\pm3$} & {$<1\times10^{23}$} & {$<1\times10^{7}$} & {1.10} & {No\tablenotemark{g}}\\
{1-385099} & {$0.9^{+0.3}_{-0.2}$} & {$<5\times10^{20}$} & {$180\pm40$} & {0.88} & {Yes}\\
{1-419607} & {$7^{+30}_{-10}$} & {$<3\times10^{23}$} & {$<4\times10^{10}$} & {1.17} & {No\tablenotemark{g}}\\
{1-456661} & {$1.1^{+0.7}_{-0.5}$} & {$<4\times10^{21}$} & {$40\pm20$} & {0.01} & {No\tablenotemark{c}}\\
{1-569225} & {$2.5^{+0.3}_{-0.2}$} & {$1.9^{+0.7}_{-0.6}\times10^{21}$} & {$10000\pm2000$} & {1.26} & {Yes}\\
{1-574506} & {$1.1^{+0.8}_{-0.6}$} & {$<1\times10^{21}$} & {$1300^{+800}_{-600}$} & {0.31} & {Yes}\\
{1-594755} & {$2.0\pm0.2$} & {$<6\times10^{20}$} & {$42000\pm4000$} & {0.56} & {Yes}\\
{1-604860} & {$1.7\pm0.1$} & {$8^{+4}_{-3}\times10^{20}$} & {$3000\pm200$} & {0.78} & {Yes}\\
{1-620993} & {$1.8\pm0.3$} & {$1\times10^{21}$} & {$470^{+80}_{-60}$} & {0.39} & {Yes}\\
\enddata
\tablenotetext{a}{The ``photon index,'' $\Gamma$ is the power-law index of the X-ray photon number spectrum, i.e. $N(E)\propto E^{-\Gamma}$ ($E$ is the photon energy and $N(E)$ is the number of photons per unit energy). It is used along with the column density, $N_{\rm H}$, are used to calculate the unabsorbed 0.3--10~keV luminosity, as described in Section~\ref{sec:agn}.\vspace{-2mm}}
\tablenotetext{b}{The hard X-ray luminosity contribution from XRBs, based on the stellar masses and SFRs from the SwiM\_all catalog file, and equation (3) of \citet{Lehmer2010}.\vspace{-2mm}}
\tablenotetext{c}{We do not conclude this object harbors an AGN due to the lack of spectroscopic evidence and the fact that the bulk of the X-ray emission in in the soft (0.3--2~keV) band. Section~\ref{ssec:obj_notes} for details.\vspace{-2mm}}
\tablenotetext{d}{The object 1-90242 is Mrk 290, which is a known AGN \citep[i.e.,][]{Bentz2015}. We cannot report the calculated $L^{\rm gal}_{\rm HX}$ from MaNGA data as the DAP cannot handle the strong AGN contribution and thus masks most of the galaxy.\vspace{-2mm}}
\tablenotetext{e}{Object 1-155975 has detectable X-ray emission but the spectrum cannot be automatically fit by the 2SXPSC software. In this case, a fixed photon index of 1.7 and Galactic N$_{\rm H}$ is assumed\vspace{-2mm}}
\tablenotetext{f}{This object resides in a large cluster, and thus the strong X-ray detection could be gas associated with the cluster. See Section~\ref{ssec:obj_notes} for details.\vspace{-2mm}}
\tablenotetext{g}{This object has weakly detected X-ray emission, but the conversion to unabsorbed luminosity results in an upper limit. We do not conclude this object harbors an AGN, due to the poorly constrained X-ray emission and conflicting spectroscopic evidence of AGN activity. See Section~\ref{ssec:obj_notes} for details.}
\end{deluxetable*}

The calculated XRB contribution presented here is limited by several effects: (1) the contamination of the observed H$\alpha$ emission from the potential AGN, and (2) the SFR is calculated within 1$R_e$. The first effect will increase the expected H$\alpha$ contribution, and thus artificially increase the fraction of hard X-ray emission explained by XRBs. While the second effect does not allow us to calculate the total X-ray emission from the galaxy, we are only interested in the nuclear emission, which is enclosed in the chosen aperture.

The 50\% light radius used for the SFR calculation is based on the SDSS $r$-band, which peaks around 6200\AA\ and encompasses the H$\alpha$ emission line. We therefore assume that the $r$-band 50\% light radius can be approximately applicable to H$\alpha$. However, even if we double the SFR, using the fact that H$\alpha$ luminosity is directly proportional to the SFR \citep[i.e.,][]{Kennicutt2012}, the XRB contribution to the hard X-ray luminosity changes by at most 50\%; the observed 0.3--10~keV luminosities are often more than a factor of 2--3 larger than the quoted $L^{\rm gal}_{\rm HX}$. Therefore, despite the large PSF of \swift/XRT, the resulting measurements can still be used to identify an AGN. We conclude that AGNs are contributing to the observed hard X-ray emission in 12 out of the 17 objects.

 \begin{figure*}[t!]
\centering
\includegraphics[width=\textwidth]{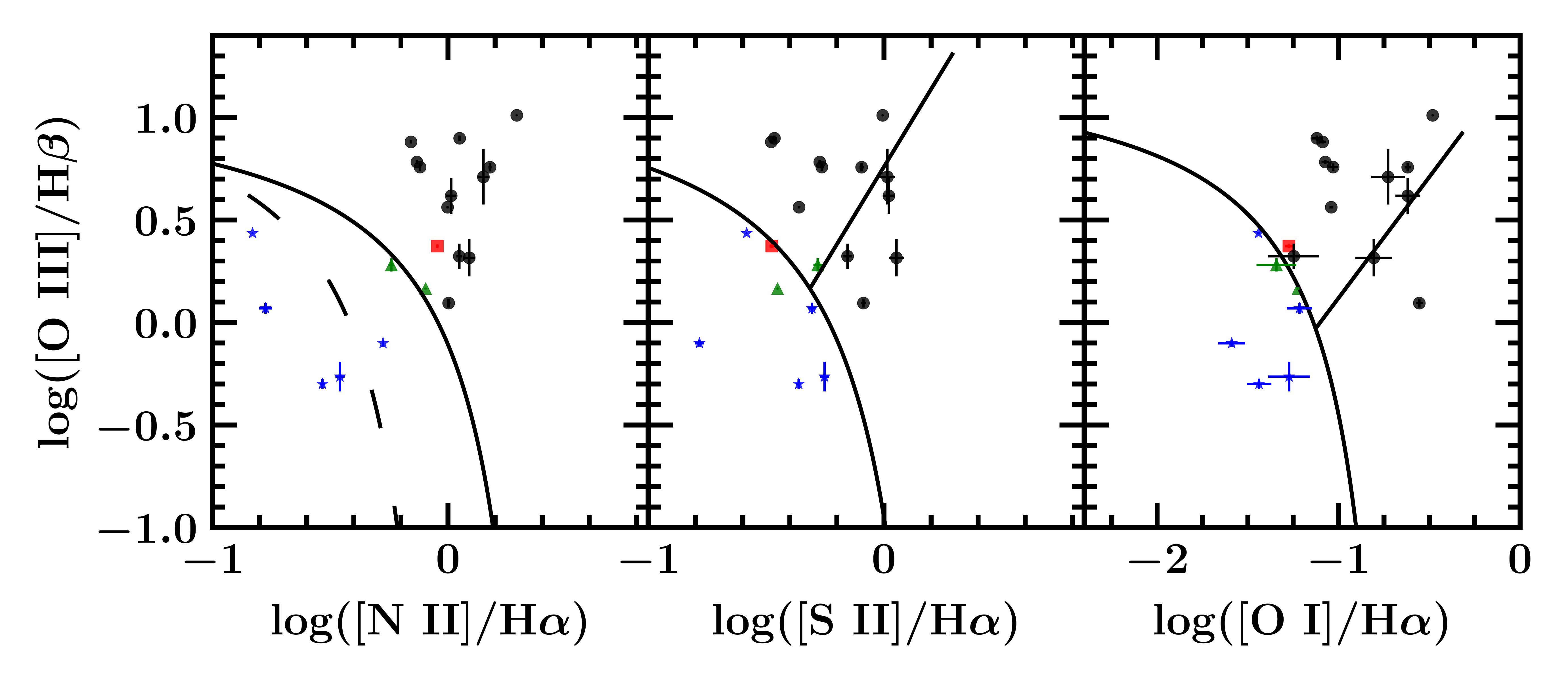}\vspace{-4mm}
\caption{The [\ion{S}{2}]/H$\alpha$, [\ion{N}{2}]/H$\alpha$ and [\ion{O}{1}]/H$\alpha$ BPT diagrams for the nuclear resolution element of the AGN candidates that have a $\textrm{S/N} > 3$ for all emission lines, before accounting for covariance. The extreme starburst lines, and Seyfert and LINER demarcation lines from \citet{Kewley2006} are plotted, along with the composite object line from \citep{Kauffmann03}. The objects that are outside the star-forming locus in all three diagrams are denoted by black filled circles, those in 2 out of the 3 diagrams are denoted by red squares, and those in 1 out of 3 diagrams are denoted by green triangles and those inside the star forming locus in all three diagrams are denoted by blue stars. 12 objects fall outside the star forming locus in all three diagrams.}\label{fig:agn_bpt} \vspace{-1mm}
\end{figure*}

In the final step, we compared the emission from the nuclear resolution element (circular aperture with a diameter of
$2\farcs 92$ to the star forming models from \citet{Kewley2006} for the [\ion{S}{2}]/H$\alpha$, [\ion{N}{2}]/H$\alpha$ and [\ion{O}{1}]/H$\alpha$ BPT diagrams, as shown in Figure~\ref{fig:agn_bpt}. We also plot the composite object line from \citet{Kauffmann03} and the LINER and Seyfert  boundary lines from \citet{Kewley2006}. We required all integrated emission line measurements to have $\textrm{S/N} > 3$ and for all spaxels within the aperture to be fit by the MaNGA DAP to be included in this test. If a galaxy meets at least one of the second or third criterion, we identify that object as an AGN.~Therefore, we conclude that the 13 objects in Table~\ref{table:xray} as well as an additional galaxy, MaNGAID 1-625513, are consistent with harboring AGN. We provide more information on galaxies with conflicting evidence for an AGN in Section~\ref{ssec:obj_notes}.

\subsection{\redtext{Notes on Individual Objects}}
\label{ssec:obj_notes}
\begin{itemize}
    \item[] \textbf{1-37155} -- This galaxy is NGC 1149, and has an upper limit on the X-ray luminosity of $L(0.3$--$10~{\rm keV}) < 2\times10^{42}$~erg~s$^{-1}$, as reported in the 1SXPSC\null. The measured emission line ratios are consistent with non-star-forming ionization mechanisms in all three BPT diagrams, but the [\ion{S}{2}]/H$\alpha$ emission line ratio fall within the LINER-region section of the diagram, which are not necessarily indicative of an AGN. The central pixel from MaNGA shows extremely weak emission lines and no AGN-like continuum. We cannot confidently conclude that NGC 1149 harbors an AGN.\vspace{-2mm}
    \item[] \textbf{1-37336} -- This galaxy has a hard to soft X-ray ratio close to zero and all three BPT diagrams are consistent with star formation. We therefore conclude that this object does not harbor an AGN.\vspace{-2mm}
       \item[] \textbf{1-43783} -- This galaxy is UGC 4056, and has an upper limit on the x-ray luminosity of $L(0.3$--$10~{\rm keV}) < 3\times10^{41}$~erg~s$^{-1}$, as reported in the 1SXPSC\null. The measured emission line ratios are consistent with non-star-forming ionization mechanisms in all three BPT diagrams, but the [\ion{S}{2}]/H$\alpha$ emission line ratios fall within the LINER-like region of the BPT diagram, which is not necessarily indicative of an AGN. Given the weak X-ray emission and ambiguous emission line ratios, we cannot confirm that UGC 4056 harbors an AGN.\vspace{-2mm}
    \item[] \textbf{1-90242} -- This galaxy is Mrk 290, a well-known AGN with strong X-ray emission. The nuclear region is not fit by the MaNGA DAP so it is not included in Figure~\ref{fig:agn_bpt}.
    \item[] \textbf{1-155975} -- This galaxy is identified as ``Star-forming'' by the SDSS collaboration, but exhibits strong X-ray emission and has Seyfert-like emission line ratios in the central MaNGA pixel. However, this signal is washed out by the $2 \farcs 92$ aperture used to construct the [\ion{S}{2}]/H$\alpha$ and [\ion{O}{1}]/H$\alpha$ BPT diagrams. Given the conflicting evidence but very strong X-ray emission, we conclude that this object may harbor a low-luminosity AGN. \vspace{-2mm}
    \item[] \textbf{1-177972} -- This object has an upper limit on its X-ray luminosity of $L(0.3$--$10~{\rm keV}) < 10^{45}$~erg~s$^{-1}$, and the spectra are not fit in the nuclear region of the IFU by the MaNGA DAP\null. This object is listed as a black hole candidate based on \textit{Hubble Space Telescope} imaging by \citet{Greene2007}, but we cannot confirm that this object harbors an AGN.\vspace{-2mm}
    \item[] \textbf{1-210784} -- This galaxy is NGC 6166C, the central cD galaxy of the Richness Class 2 cluster Abell 2199.  The observed X-rays in the 0.3--10~keV band could therefore be from to the system's intracluster medium.  Additionally, the nuclear [\ion{O}{1}] emission is very weak, with a S/N $< 3$. In combination with the weak X-ray emission and the very weak [\ion{O}{1}] emission, we cannot confidently conclude that NGC 6166C harbors an AGN.\vspace{-2mm}
    \item[] \textbf{1-211023} -- This galaxy has an upper limit on the X-ray luminosity of $L(0.3$--$10~{\rm keV}) < 4\times10^{43}$~erg~s$^{-1}$, as reposrted in the 1SXPSC\null. The object's measured emission line ratios are consistent with non-star-forming ionization mechanisms in all three BPT diagrams, though the [\ion{O}{1}]/H$\alpha$ and [\ion{S}{2}]/H$\alpha$ ratios are in the LINER region of the corresponding diagrams.  We therefore do not conclude that this object harbors an AGN.\vspace{-2mm}
    \item[] \textbf{1-258876} -- The X-ray luminosity of this galaxy is  $L(0.3$--$10~{\rm keV}) < 9\times10^{40}$~erg~s$^{-1}$, as reported in the 1SXPSC\null. The measured emission line ratios are again consistent with non-star-forming ionization mechanisms, but the [\ion{S}{2}]/H$\alpha$ emission line ratio fall within the LINER region of the BPT diagram, which is not necessarily indicative of an AGN. In addition, the [O I] emission is not measured in all pixels within the central $2\farcs 92$ aperture, which prevents a robust AGN identification. We therefore cannot conclude that this object harbors an AGN.\vspace{-2mm}
    \item[] \textbf{1-269227} -- This galaxy has strong X-ray emission in the 0.3--10~keV band, but the H$\beta$ emission line flux has a S/N $< 3$ in the central aperture, making the analysis in all BPT diagrams untrustworthy. However, we use the X-ray emission to conclude that the galaxy harbors an AGN.\vspace{-2mm}
    \item[] \textbf{1-281439} -- This galaxy  has an upper limit on the X-ray luminosity of $L(0.3$--$10~{\rm keV}) < 2\times10^{43}$~erg~s$^{-1}$, as reported in the 1SXPSC\null. The galaxy shows Seyfert or AGN-like emission line ratios in the [\ion{O}{1}]/H$\alpha$ and [\ion{N}{2}]/H$\alpha$ diagrams, but its [\ion{S}{2}]/H$\alpha$ ratio is consistent with excitation by star formation. Given the conflicting evidence and non-detection in the X-ray, we cannot confirm the presence of an AGN in this galaxy.\vspace{-2mm}
    \item[] \textbf{1-317315} -- This galaxy has poorly-constrained X-ray emission as measured by XRT\null. While there is ``observed'' emission, the unabsorbed X-ray flux is statistically consistent with zero. The object also displays Seyfert or AGN-like  [\ion{N}{2}]/H$\alpha$ ratios, but falls within the LINER-like region of the [\ion{S}{2}]/H$\alpha$ and [\ion{O}{1}]/H$\alpha$ diagrams. The central pixel of the MaNGA data shows very weak emission lines and no AGN-like continuum. We cannot confidently conclude that this object harbors an AGN.\vspace{-2mm}
   \item[] \textbf{1-456661} --This galaxy is not identified as an AGN by its emission line, but has X-ray emission detected by XRT\null. However, the galaxy resides in the Coma cluster, so the X-ray emission could be a result of the hot gas from the cluster and not necessarily indicative of an AGN.~We therefore conclude the object does not harbor an AGN.\vspace{-2mm}
      \item[] \textbf{1-419607} -- This galaxy shows no spectroscopic signatures of an AGN, and the X-ray emission is poorly constrained. While X-rays are detected, the unabsorbed X-ray emission is still statistically consistent with zero. Therefore we do not conclude that this object harbors an AGN.\vspace{-2mm}
   \item[] \textbf{1-569225} --This galaxy is identified as Star-forming by the SDSS collaboration, but exhibits strong X-ray emission. The measured emission line ratios are consistent with non-star-forming ionization mechanisms in all three BPT diagrams, but the [\ion{O}{1}]/H$\alpha$ and [\ion{S}{2}]/H$\alpha$ emission line ratios fall within the LINER-like region of the diagram, which are not necessarily indicative of an AGN. Given the strong X-ray detection, we conclude this object harbors an AGN.\vspace{-2mm}
   \item[] \textbf{1-625513} --This galaxy is IC 4227, and the upper limit on its x-ray luminosity, $L(0.3$--$10~{\rm keV}) < 2\times10^{41}$~erg~s$^{-1}$ as reported in the 1SXPSC does not rule out the presence of an AGN. Additionally, the measured emission line ratios are consistent with AGN or Seyfert-like ionization mechanisms in all three BPT diagrams, and the central MaNGA pixel shows strong narrow emission lines consistent with a Seyfert 2 spectrum. We conclude that IC 4227 harbors an AGN.\vspace{-2mm}
\end{itemize}

\section{Summary}\label{sec3:conclusion}
In this paper, we present the SwiM value added catalog, which consists of 150 galaxies in the MaNGA main sample that have also been observed by \textit{Swift}/UVOT in both the uvw2 and uvw1 filters. In this dataset, we provide the integrated photometry in the UVOT uvw2, uvm2, and uvw1 filters measured consistently with the SDSS and \galex\ photometry provided by the NSA, along with selection weights and scaling factors for correcting back to a volume-limited sample. We also present resolution- and sampling-matched SDSS and Swift images, along with matching maps for emission-line and spectral indices based on MaNGA spectra. Errors have been propagated taking into account of the covariance throughout the reduction. We also present the correlation between pixels in the resulting maps. All the images and maps have a final resolution of $2\farcs 92$ FWHM and are sampled on 1\arcsec\ pixels. AGNs have been identified in the sample using both optical line ratio diagnostics and X-ray data. These data will be very useful for studying spatially-resolved dust attenuation and star formation histories within galaxies. We make these data publicly available on the SDSS website as a VAC. 

\acknowledgements
We thank the referee for their thoughtful comments. We thank Lea Hagen, Michael Siegel and David Molina for their help with some of the technical aspects of this work. This work was supported by NASA through grant number 80NSSC20K0436 (ADAP). This work was supported by funding from the Alfred P. Sloan Foundation's Minority Ph.D. (MPHD) Program, awarded to MM in 2014--15. This work made use of data supplied by the UK Swift Science Data Centre at the University of Leicester. This research has made use of data and/or software provided by the High Energy Astrophysics Science Archive Research Center (HEASARC), which is a service of the Astrophysics Science Division at NASA/GSFC and the High Energy Astrophysics Division of the Smithsonian Astrophysical Observatory. This research made use of Astropy, a community-developed core Python package for Astronomy \citep{Whelan2018}.  This research has made use of the NASA/IPAC Extragalactic Database (NED), which is operated by the Jet Propulsion Laboratory, California Institute of Technology, under contract with the National Aeronautics and Space Administration. This research has made use of data obtained from the 3XMM XMM-Newton serendipitous source catalog compiled by the 10 institutes of the XMM-Newton Survey Science Centre selected by ESA. This publication makes use of data products from the Wide-field Infrared Survey Explorer, which is a joint project of the University of California, Los Angeles, and the Jet Propulsion Laboratory/California Institute of Technology, and NEOWISE, which is a project of the Jet Propulsion Laboratory/California Institute of Technology. WISE and NEOWISE are funded by the National Aeronautics and Space Administration. This publication makes use of data products from the Two Micron All Sky Survey, which is a joint project of the University of Massachusetts and the Infrared Processing and Analysis Center/California Institute of Technology, funded by the National Aeronautics and Space Administration and the National Science Foundation. This research has made use of the NASA/IPAC Infrared Science Archive, which is operated by the Jet Propulsion Laboratory, California Institute of Technology, under contract with the National Aeronautics and Space Administration. The Institute for Gravitation and the Cosmos is supported by the Eberly College of Science and the Office of the Senior Vice President for Research at the Pennsylvania State University. Funding for the Sloan Digital Sky Survey IV has been provided by the Alfred P. Sloan Foundation, the U.S. Department of Energy Office of Science, and the Participating Institutions. SDSS-IV acknowledges support and resources from the Center for High-Performance Computing at the University of Utah. The SDSS web site is www.sdss.org. 

SDSS-IV is managed by the Astrophysical Research Consortium for the Participating Institutions of the SDSS Collaboration including the Brazilian Participation Group, the Carnegie Institution for Science, Carnegie Mellon University, the Chilean Participation Group, the French Participation Group, Harvard-Smithsonian Center for Astrophysics, Instituto de Astrof\'isica de Canarias, The Johns Hopkins University, Kavli Institute for the Physics and Mathematics of the Universe (IPMU) / University of Tokyo, the Korean Participation Group, Lawrence Berkeley National Laboratory, Leibniz Institut f\"ur Astrophysik Potsdam (AIP),  Max-Planck-Institut f\"ur Astronomie (MPIA Heidelberg), Max-Planck-Institut f\"ur Astrophysik (MPA Garching), Max-Planck-Institut f\"ur Extraterrestrische Physik (MPE), National Astronomical Observatories of China, New Mexico State University, New York University, University of Notre Dame, Observat\'ario Nacional / MCTI, The Ohio State University, Pennsylvania State University, Shanghai Astronomical Observatory, United Kingdom Participation Group,Universidad Nacional Aut\'onoma de M\'exico, University of Arizona, University of Colorado Boulder, University of Oxford, University of Portsmouth, University of Utah, University of Virginia, University of Washington, University of Wisconsin, Vanderbilt University, and Yale University.

\clearpage

\appendix

\section{S\MakeLowercase{wi}M VAC Data Model}
\subsection{Catalog Data Model}\label{app:vac_mod}
This appendix provides the SwiM VAC data model for the SwiM\_all catalog file. The catalog file holds basic properties of the galaxies included in the \swim, as well as the integrated \galex, \swift/UVOT, and SDSS photometry. The names and contents of each extension in this file is given in Table~\ref{table:catalog_dm}.

\begin{deluxetable}{lrl}[H]
  \tablecaption{SwiM Catalog Data Model \label{table:catalog_dm}}
\tabletypesize{\footnotesize}
\setlength{\tabcolsep}{6pt}
\renewcommand{\arraystretch}{1.1}
\tablewidth{0pt}
\tablehead{{Column} & {Units} & {Description}}
\startdata
{\texttt{MANGAID}} & \dots & MaNGA ID for the object (e.g 1-210754)\\
\texttt{PLATE} & \dots & Plate ID for the object\\
\texttt{IFUDSGN} & \dots & IFU design ID for the object (e.g. 12701)\\
\texttt{MNGTARG1} & \dots & MANGA\_TARGET1 maskbit for galaxy target catalog\\
\texttt{MNGTARG3} &\dots & MANGA\_TARGET3 maskbit for galaxy target catalog\\
\texttt{NAME} & \dots & Galaxy Name\\
\texttt{SDSS\_CLASS} & \dots & SDSS DR15 object classification\\
\texttt{EBV} & \dots & $E(B-V)$ value from \cite{Schlegel1998} dust map for this galaxy\\
\texttt{RA} & {deg} & Right-ascension of the galaxy center in J2000\\
\texttt{DEC} & {deg} & Declination of the galaxy center in J2000\\
\texttt{NSA\_ELPETRO\_PHI} & {deg} & Position angle (east of north) used for elliptical apertures\\
\texttt{NSA\_ELPETRO\_TH50\_R} & {arcsec} & Elliptical Petrosian 50\% light radius (semi-major axis) in SDSS r-band\\
\texttt{NSA\_ELPETRO\_THETA} & {arcsec} & Elliptical Petrosian radius (semi-major axis) in SDSS r-band\\
\texttt{NSA\_ELPETRO\_BA} & \dots & Axis ratio used for elliptical apertures\\
\texttt{NSA\_ELPETRO\_MASS} & $h^{-2}$ solar masses & Stellar mass from K-correction fit for elliptical Petrosian fluxes\\
\texttt{NSA\_Z} & \dots & Heliocentric redshift from the NASA-Sloan Atlas\\
\multirow{2}{*}{}\texttt{NSA\_ELPETRO\_FLUX} & nanomaggies & Elliptical SDSS-style Petrosian flux the \galex\ and SDSS filters in \\ & & bands [FNugriz] (using r-band aperture)\\
\texttt{NSA\_ELPETRO\_FLUX\_IVAR} & nanomaggies$^{-2}$ & Inverse variance of \texttt{NSA\_ELPETRO\_FLUX} [FNugriz]\\
\multirow{2}{*}{}\texttt{SWIFT\_ELPETRO\_FLUX} & nanomaggies & Elliptical SDSS-style Petrosian flux in bands [uvw2, uvm2, uvw1]\\ & & (aperture corrected using r-band aperture)\\
\texttt{SWIFT\_ELPETRO\_FLUX\_IVAR} & nanomaggies$^{-2}$ & Inverse variance for \multirow{2}{*}{}\texttt{SWIFT\_ELPETRO\_FLUX} [uvw2, uvm2, uvw1]; \\ & & if there is no uvm2 measurement that element is $-1$\\
\multirow{3}{*}{}\texttt{SWIFT\_EXPOSURE} & sec & Exposure times for Swift/UVOT bands [uvw2, uvm2, uvw1]; \\ & & if there is no uvm2 measurement that element is $-1$\\
\multirow{3}{*}{}\texttt{APERCORR} & \dots & Aperture correction factor $f_a/f_b$ for Swift/UVOT bands [uvw2, uvm2, uvw1];\\ & & $f_a$ and $f_b$ are the r-band integrated fluxes (of the mock galaxy) before and after \\ & & the Swift/UVOT PSF convolution (see Section~\ref{sec:int_phot})\\
\texttt{SFR\_1RE} & \dots & Dust corrected log(SFR/$1\;M_\odot\;$yr$^{-1}$) using H$\alpha$ flux within 1 effective radius reported\\ & & in MaNGA DAP \citep[see][and Section~\ref{ssec:sm_comp} of this work]{Westfall2019}\\
\texttt{SCALING\_FACTOR} & \dots & Scaling factors that represent the number of objects in the SwiM catalog \\ & &  divided by the number in MaNGA main sample in each bin (see Section~\ref{sec:volwgt})\\
\texttt{SCALING\_FACTOR\_ERR} & \dots & 1$\sigma$ uncertainty for \texttt{SCALING\_FACTOR}s\\
\texttt{ESWEIGHT} & \dots & Volume weights from \citet{Wake2017} for Primary+ and full secondary sample\\
\enddata
\end{deluxetable}

\clearpage

\subsection{Map HDU Data Models}\label{app:hdu_mod}
In this appendix we present the data model for the SwiM VAC map files. The maps files contain the spatially-matched MaNGA IFU maps, \swift/UVOT and SDSS photometry. The map data files have 17 total HDUs, with 5 main groups: D$_n$(4000) (HDU 0), spectral indices (HDU 1--8), emission line flux and equivalent width (HDU 9--14), \swift\ and SDSS photometry (HDU 15--16), and the raw \swift\ information (HDU 17). The HDU format for each group is described below, including notes on how to use the maps. The names and descriptions of the header data units (HDUs) are given in Table~\ref{table:maps_dm}, while the formatting of the HDUs are described in Tables~\ref{table:d4000_dm} through~\ref{table:rawswft_dm}.  

All MaNGA maps and UVOT images have masks, where science-ready pixels are indicated by 0 and 1 otherwise. The MaNGA masks are based on those in DR15, but has been simplified to a 0 or 1 mask given the analysis presented in this work. The masks for the UVOT images only affect two objects as discussed in Section~\ref{sec:sw_man_reduce}.

\begin{deluxetable}{clcrl}[H]
  \tablecaption{SwiM Maps HDU Descriptions\label{table:maps_dm}}
\tabletypesize{\footnotesize}
\setlength{\tabcolsep}{6pt}
\renewcommand{\arraystretch}{1.1}
\tablewidth{0pt}
\tablehead{
{Index} &{Name} & {Channels} &{Units} & {Description}}
\startdata
\multirow{2}{*}{} 0 & \texttt{Dn4000} & {5} & {erg s$^{-1}$ cm$^{-2}$ Hz$^{-1}$ arcsec$^{-2}$} & Maps required to calculate the D$_n$(4000) measurements\\ & & & & and its uncertainty\\
{1} & \texttt{SPECINDX\_FLUX} & {43} & {erg s$^{-1}$ cm$^{-2}$ arcsec$^{-2}$} & Spectral index flux maps ($F_{I}$ in equation \ref{eqn:E4.1})\\
{2} & \texttt{SPECINDX\_CONT} & {43} & {erg s$^{-1}$ cm$^{-2}$ \AA$^{-1}$ arcsec$^{-2}$} & Spectral index continuum maps ($F_{\rm C0}$ in equation \ref{eqn:E3})\\
{3} & \texttt{SPECINDX\_FLUX\_SIGMA} & {43} & erg s$^{-1}$ cm$^{-2}$ arcsec$^{-2}$ & 1$\sigma$ uncertainties for \texttt{SPECINDX\_FLUX}\\
{4} & \texttt{SPECINDX\_CONT\_SIGMA} & {43} & {erg s$^{-1}$ cm$^{-2}$ \AA$^{-1}$ arcsec$^{-2}$} & 1$\sigma$ uncertainties for \texttt{SPECINDX\_CONT}\\
\multirow{2}{*}{}{5} & \texttt{SPECINDX\_MASK} & {43} & \dots & Masks for \texttt{SPECINDX\_FLUX}, \texttt{SPECINDX\_FLUX\_SIGMA}, \\ & & & &  \texttt{SPECINDX\_CONT} and \texttt{SPECINDX\_CONT\_SIGMA} \\
{6} & \texttt{COMBINED\_DISP} & {43} & {km s$^{-1}$} & Flux-weighted combined dispersion maps\\
{7} & \texttt{COMBINED\_DISP\_SIGMA} & {43} & {km s$^{-1}$} & 1$\sigma$ uncertainties for \texttt{COMBINED\_DISP}\\
{8} & \texttt{COMBINED\_DIPS\_MASK} & {43} & \dots & Masks for \texttt{COMBINED\_DISP} and \texttt{COMBINED\_DISP\_SIGMA}\\
{9} & \texttt{ELINE\_FLUX} & {22} & {10$^{-17}$ erg s$^{-1}$ cm$^{-2}$ arcsec$^{-2}$} & Gaussian-fitted emission line flux maps based on MPL-7 DAP\\
{10} & \texttt{ELINE\_FLUX\_SIGMA} & {22} & {10$^{-17}$ erg s$^{-1}$ cm$^{-2}$ arcsec$^{-2}$} & 1$\sigma$ uncertainties for \texttt{ELINE\_FLUX}\\
{11} & \texttt{ELINE\_FLUX\_MASK} & {22} & \dots & Masks for \texttt{ELINE\_FLUX} and \texttt{ELINE\_FLUX\_SIGMA}\\
{12} & \texttt{ELINE\_EW} & {22} & {\AA} & Gaussian-fitted equivalent width maps based on MPL-7 DAP\\
{13} & \texttt{ELINE\_EW\_SIGMA} & {22} & {\AA} & 1$\sigma$ uncertainties for the \texttt{ELINE\_EW\_SIGMA}\\
{14} & \texttt{ELINE\_EW\_MASK} & {22} & \dots & Masks for \texttt{ELINE\_EW} and \texttt{ELINE\_EW\_SIGMA}\\
\multirow{2}{*}{}{15} & \texttt{SWIFT/SDSS} & {8} & {nanomaggies} & \textit{Swift}/UVOT and SDSS sky-subtracted images\\ & & & & [uvw2,uvw1,uvm2,u,g,r,i,z]\\
{16} & \texttt{SWIFT/SDSS\_SIGMA} & {8} & {nanomaggies} & 1$\sigma$ uncertainties for \texttt{SWIFT/SDSS}\\
\multirow{2}{*}{}{17} & \texttt{SWIFT\_UVOT} & {12} & \dots &  Swift/UVOT non-sky-subtracted counts, exposure, \\ & & & & counts error and mask maps [uvw2, uvw1, uvm2].
\enddata
\end{deluxetable}

\clearpage

\textbf{HDU 0: D4000} -- This HDU contains the maps necessary to calculate D$_n$(4000) measurements and their uncertainties. The data are a 3--D array with the third dimension having a size of 5 corresponding to the two data channels, their uncertainties and the mask. To calculate D$_n$(4000), use the relation $\textrm{D}_n(4000) = f_{\nu,\rm{red}}/f_{\nu,\rm{blue}}$, and its uncertainty as
\begin{equation}
\sigma_{\textrm{D}_n(4000)}=\textrm{D}_n(4000)\sqrt{\left(\frac{\sigma_{f_{\nu,\rm{red}}}}{{f_{\nu,\rm{red}}}}\right)^{\!\!2}+\left(\frac{\sigma_{f_{\nu,\rm{blue}}}}{{f_{\nu,\rm{blue}}}}\right)^{\!\!2}}. 
\end{equation}
Covariance has been properly taken into account in our data processing. 
The final uncertainty must be multiplied by 1.4 to account for calibration errors described in \citet{Westfall2019}. See Section~\ref{sec:manerr} for more information. 

All maps have the units erg s$^{-1}$ cm$^{-2}$ Hz$^{-1}$ arcsec$^{-2}$, except for the mask, which uses 0 to indicate a science-ready pixels and 1 otherwise. The structure of the HDU is given in Table~\ref{table:d4000_dm}.

\begin{deluxetable}{cll}[H]
  \tablecaption{HDU0: D4000 Channel Description\label{table:d4000_dm}}
\tabletypesize{\footnotesize}
\setlength{\tabcolsep}{6pt}
\renewcommand{\arraystretch}{1.1}
\tablewidth{0pt}
\tablehead{
{Channel} &{Name} & {Description}}
\startdata
{0} & \texttt{Fnu Red} & {Flux density per unit wavelength in the red window}\\
{1} & \texttt{Fnu Blue} & {Flux density per unit wavelength in the blue window}\\
{2} & \texttt{Sigma Red} & {Uncertainty in flux density in the red window}\\
{3} & \texttt{Sigma Blue} & {Uncertainty in flux density in the blue window}\\
{4} & \texttt{Mask}  & D$_n$(4000) mask\\
\enddata
\end{deluxetable}

\clearpage
\textbf{HDU 1-8: SPECINDX} -- These HDUs contain the information needed to calculate the spectral indices included in this VAC \citep[the first 43 listed in][]{Westfall2019}. We put D$_n$(4000) spectral index in HDU 0 because of its different definition and units. The spectral indices can be calculated using Equation~\ref{eqn:E4.1}.

This relation is described in more detail in Section~\ref{ssec:lickidx}. HDUs 1 and 3 contain the flux and the uncertainty measurements for the index flux $F_I$ in units of erg s$^{-1}$ cm$^{-2}$ arcsec$^{-2}$, while HDUs 2 and 4 contain the same information for the continuum flux density $f_{C0}$ in units of erg s$^{-1}$ cm$^{-2}\;$\AA$^{-1}$ arcsec$^{-2}$. HDU 5 is the mask for the spectral index maps, where 0 denotes science-ready pixels, and 1 denotes otherwise. 

Each HDU contains a 3D array with the third dimension having a size of 43 (channels) corresponding to the 43 indices. The channel-to-index mapping is provided in the header and in Table~\ref{table:specindx_dm}. Here we also include the $\Delta \lambda$ for each index which is needed to compute the final indices using Equation~\ref{eqn:E4.1}.  The index bandpasses are identical to that given in Table 4 of \citet{Westfall2019}. 
 
 HDUs 6--8 contain the flux weighted combined stellar and instrumental dispersion maps, its uncertainty, and mask. The data in HDUs 6 and 7 are in units of km s$^{-1}$, while the masks in HDU 8 have the same definitions as that of HDU 5. These HDUs also have 43 channels corresponding to the 43 indices as given in their header and in  Table~\ref{table:specindx_dm}.

\begin{deluxetable}{cll}[H]
  \tablecaption{HDU 1--8: Spectral Index Channels Description\label{table:specindx_dm}}
\tabletypesize{\footnotesize}
\setlength{\tabcolsep}{20pt}
\renewcommand{\arraystretch}{1.1}
\tablehead{
{Channel} &{Name} & {$\Delta\lambda$\tablenotemark{a} (\AA)}}
\startdata
{0} &  \texttt{CN1} & {35}\\
{1} &  \texttt{CN2} & {35}\\
{2} &  \texttt{Ca4227} & {12.5}\\
{3} &  \texttt{G4300} & {35}\\
{4} &  \texttt{Fe4383} & {51.25}\\
{5} &  \texttt{Ca4455} & {22.5}\\
{6} &  \texttt{Fe4531} & {45}\\
{7} &  \texttt{C24668} & {86.25}\\
{8} &  \texttt{H$\beta$} & {28.75}\\
{9} &  \texttt{Fe5015} & {76.25}\\
{10} & \texttt{Mg1} & {65}\\
{11} & \texttt{Mg2} & {42.5}\\
{12} & \texttt{Mgb} & {32.5}\\
{13} & \texttt{Fe5270} & {40}\\
{14} & \texttt{Fe5335} & {40}\\
{15} & \texttt{Fe5406} & {27.5}\\
{16} & \texttt{Fe5709} & {23.75}\\
{17} & \texttt{Fe5782} & {20}\\
{18} & \texttt{NaD} & {32.5}\\
{19} & \texttt{TiO1} & {57.5}\\
{20} & \texttt{TiO2} & {82.5}\\
{21} & \texttt{H$\delta_\textrm{A}$} & {38.75}\\
{22} & \texttt{H$\gamma_\textrm{A}$} & {43.75}\\
{23} & \texttt{H$\delta_\textrm{F}$} & {21.25}\\
{24} & \texttt{H$\gamma_\textrm{F}$} & {21}\\
{25} & \texttt{CaHK} & {104}\\
{26} & \texttt{CaII1} & {29}\\
{27} & \texttt{CaII2} & {40}\\
{28} & \texttt{CaII3} & {40}\\
{29} & \texttt{Pa17} & {13}\\
{30} & \texttt{Pa14} & {42}\\
{31} & \texttt{Pa12} & {42}\\
{32} & \texttt{MgICvD} & {55}\\
{33} & \texttt{NaICvD} & {28}\\
{34} & \texttt{MgIIR} & {15}\\
{35} & \texttt{FeHCvD} & {30}\\
{36} & \texttt{NaI} & {65.625}\\
{37} & \texttt{bTiO} & {41.5}\\
{38} & \texttt{aTiO} & {155}\\
{39} & \texttt{CaH1} & {44.25}\\
{40} & \texttt{CaH2} & {125}\\
{41} & \texttt{NaISDSS} & {20}\\
{42} & \texttt{TiO2SDSS} & {82.5}\\
\enddata
\tablenotetext{a}{The $\Delta\lambda$ presented here is the width of the index band.}
\end{deluxetable}

\clearpage

\textbf{HDU 9--14: ELINE\_FLUX and ELINE\_EW} -- These HDUs contain the emission line flux maps and EW maps, and their associated uncertainties. The emission line fluxes come from the Gaussian-fitted measurements from the MPL-7 DAP. Again each HDU contains a 3D array with the third dimension corresponding to different emission line channels. The channel-to-line mapping can be found in the header and in Table~\ref{table:flux_dm}. 

HDUs 9 and 10 contain the measured flux and uncertainty in units of $10^{-17}$ erg s$^{-1}$ cm$^{-2}$ arcsec$^{-2}$, while HDUs 12 and 13 contain the EW information in units of \AA. HDUs 11 and 14 contain the masks for flux and EW, respectively, defined so that 0 denotes science-ready pixels and 1 denotes otherwise. 

\begin{deluxetable}{cll}[H]
  \tablecaption{HDU 9--14: Emission Line Channels Description\label{table:flux_dm}}
\tabletypesize{\footnotesize}
\setlength{\tabcolsep}{20pt}
\renewcommand{\arraystretch}{1.1}
\tablehead{
{Channel} &{Ion} & {$\lambda_{\textrm{rest}}$ (\AA)}\tablenotemark{a}}
\startdata
{0} &  \texttt{[\ion{O}{2}]} & {3727.092}\\
{1} &  \texttt{[\ion{O}{2}]} & {3729.875}\\
{2} &  \texttt{H$\theta$} & {3798.9826}\\
{3} &  \texttt{H$\eta$} & {3836.4790}\\
{4} &  \texttt{[\ion{Ne}{3}]} & {3869.86}\\
{5} &  \texttt{H$\zeta$} & {3890.1576}\\
{6} &  \texttt{[\ion{Ne}{3}]} & {3968.59}\\
{7} &  \texttt{H$\epsilon$} & {3971.2020}\\
{8} &  \texttt{H$\delta$} & {4102.8991}\\
{9} &  \texttt{H$\gamma$} & {4341.691}\\
{10} & \texttt{[\ion{He}{2}]} & {4687.015}\\
{11} & \texttt{H$\beta$} & {4862.691}\\
{12} & \texttt{[\ion{O}{3}]} & {4960.295}\\
{13} & \texttt{[\ion{O}{3}]} & {5008.240}\\
{14} & \texttt{[\ion{He}{1}]} & {5877.243}\\
{15} & \texttt{[\ion{O}{1}]} & {6302.046}\\
{16} & \texttt{[\ion{O}{1}]} & {6365.535}\\
{17} & \texttt{[\ion{N}{2}]} & {6549.86}\\
{18} & \texttt{H$\alpha$} & {6564.632}\\
{19} & \texttt{[\ion{N}{2}]} & {6585.271}\\
{20} & \texttt{[\ion{S}{2}]} & {6718.294}\\
{21} & \texttt{[\ion{S}{2}]} & {6732.674}\\
\enddata
\tablenotetext{a}{Vacuum rest wavelengths from the National Institute of Standards and Technology (NIST) and are used by the MaNGA DAP.}
\end{deluxetable}

\clearpage
\textbf{HDU 15 --16: SWIFT/SDSS} -- HDU15 contains the sky-subtracted NUV images from \swift\ and optical images from SDSS, and HDU16 contains their corresponding uncertainty images. All images are provided in units of nanomaggies. To convert these maps to the AB magnitude ($m$) system, use $m = 22.5 - 2.5\log_{\textrm{10}}(f/\textrm{nanomaggie})$. To convert to $\mu$Jy use 1 $\textrm{nanomaggie}= 3.631$~$\mu$Jy. 

For 5 out of the 150 galaxies, there are bad pixels that report incorrect exposure times (usually 0 or NaN) within the map's FoV. In this file, the values are stored as $-$inf or NaN, which will cause errors in photometric measurements if not masked. We provide masks for all Swift images in HDU 17. We strongly recommend users always use the masks from HDU 17 when working with \swift\ images. If there are no bad pixels, then the mask will not change the image. For SDSS, masked pixels have an uncertainty of 0. 

In these two HDUs, the data are also given in 3D arrays with the third dimension corresponding to different filters. The correspondence are given in the header and in Table~\ref{table:phot_dm}.

\begin{deluxetable}{clc}[H]
  \tablecaption{HDU 15--16: Photometry Channel Description\label{table:phot_dm}}
\tabletypesize{\footnotesize}
\setlength{\tabcolsep}{6pt}
\renewcommand{\arraystretch}{1.1}
\tablewidth{0pt}
\tablehead{
{Channel} &{Name} & {Central Wavelength (\AA)}}
\startdata
{0} & \texttt{uvw2} & {1928}\\
{1} & \texttt{uvw1} & {2600}\\
{2} & \texttt{uvm2} & {2246}\\
{3} & \texttt{SDSS u} & {3543}\\
{4} & \texttt{SDSS g}  & {4770}\\
{5} & \texttt{SDSS r} & {6231}\\
{6} & \texttt{SDSS i} & {7625}\\
{7} & \texttt{SDSS z} & {9134}\\
\enddata
\end{deluxetable}

\clearpage
\textbf{HDU 17: SWIFT\_UVOT} -- This HDU contains the \swift/UVOT counts, uncertainty, exposure and mask maps for all three NUV filters. The masks have a value of 0 for science-ready pixels and 1 otherwise. Unlike HDU 15 and 16, these images are not sky-subtracted. However the calculated sky counts are provided in the header under the keywords \texttt{SKY\_W1}, \texttt{SKY\_M2} and \texttt{SKY\_W2}. The AB magnitude system zero points of the filters and $f_{\lambda}$ conversion factors are also provided in the header as \texttt{ABZP\_}$\ast$ and \texttt{FLAMBDA\_}$\ast$, where the $\ast$ represents the desired filter. 
The structure of this HDU is given in Table~\ref{table:rawswft_dm}.
\begin{deluxetable}{cll}[H]
  \tablecaption{HDU 17: Swift/UVOT Channel Description\label{table:rawswft_dm}}
\tabletypesize{\footnotesize}
\setlength{\tabcolsep}{6pt}
\renewcommand{\arraystretch}{1.1}
\tablewidth{0pt}
\tablehead{
{Channel} &{Name} & {Description}}
\startdata
{0} & \texttt{uvw2 Counts} & {Fully reduced, non-sky subtracted uvw2 counts}\\
{1} & \texttt{uvw1 Counts} & {Fully reduced, non-sky subtracted uvw1 counts}\\
{2} & \texttt{uvm2 Counts} & {Fully reduced, non-sky subtracted uvm2 counts}\\
{3} & \texttt{uvw2 Counts Err} & {Uncertainty associated with uvw2 counts}\\
{4} & \texttt{uvw1 Counts Err} & {Uncertainty associated with uvw1 counts}\\
{5} & \texttt{uvm2 Counts Err} & {Uncertainty associated with uvm2 counts}\\
{6} & \texttt{uvw2 Exposure} & {Exposure map for uvw2 image}\\
{7} & \texttt{uvw1 Exposure} & {Exposure map for uvw1 image}\\
{8} & \texttt{uvm2 Exposure} & {Exposure map for uvm2 image}\\
{9} & \texttt{uvw2 Mask} & {Mask for uvw2 image}\\
{10} & \texttt{uvw1 Mask} & {Mask for uvw1 image}\\
{11} & \texttt{uvm2 Mask} & {Mask for uvm2 image}\\
\enddata
\end{deluxetable}
\clearpage

\bibliographystyle{apj}
\bibliography{all.041018.bib}

\end{document}